\documentclass[12pt,preprint]{aastex}
\usepackage{epstopdf}
\usepackage{bm}
\def\Meszaros{M\'esz\'aros~}
\usepackage{epsfig}
\slugcomment{ }

\begin{document}

\title {On the High Energy Emission of the Short GRB 090510}
\author {Hao-Ning He\altaffilmark{1,2,4,5}, Xue-Feng Wu\altaffilmark{2,3,4},
Kenji Toma\altaffilmark{2,4}, Xiang-Yu Wang\altaffilmark{1,5} and
Peter \Meszaros\altaffilmark{2,4}}
\altaffiltext{1}{Department of Astronomy, Nanjing University, Nanjing 210093, China}
\altaffiltext{2}{Department of Astronomy \& Astrophysics, Department of Physics,
    Pennsylvania State University, 525 Davey Lab, University Park, PA 16802}
\altaffiltext{3}{Purple Mountain Observatory, Chinese Academy of Sciences,
   Nanjing 210008, China}
\altaffiltext{4}{Center for Particle Astrophysics,
    Pennsylvania State University, 104 Davey Lab, University Park, PA 16802}
\altaffiltext{5}{Key Laboratory of Modern Astronomy and
Astrophysics (Nanjing University), Ministry of Education, Nanjing 210093, China}
\begin{abstract}

Long-lived high-energy ($>100 {\rm MeV}$) emission, a common
feature of most Fermi-LAT detected gamma-ray burst, is detected up
to $\sim10^2$ s in the short GRB 090510. We study the origin
of this long-lived high-energy emission, using
broad-band observations including X-ray and optical data. We confirm
that the late $> 100$ MeV, X-ray and optical emission can be
naturally explained via synchrotron emission from an adiabatic
forward shock propagating into a homogeneous ambient medium with low
number density. The Klein-Nishina effects are found to be
significant, and effects due to jet spreading and magnetic field
amplification in the shock appear to be required. Under the
constraints from the low-energy observations, the adiabatic forward shock
synchrotron emission is consistent with the later-time ($t \gtrsim
2{\rm s}$) high-energy emission, but falls below the early-time
($t<2 {\rm s}$) high energy emission. 
Thus we argue that an extra high energy component is needed at early times.
A standard reverse shock origin is found to be inconsistent with this extra component. Therefore,
we attribute the early part of the high-energy emission ($t \lesssim 2 \rm s$) 
to the prompt component, and the long-lived high energy emission ($t\ga 2\rm s$)
to the adiabatic forward shock synchrotron afterglow radiation. This avoids the
requirement for an extremely high initial Lorentz factor.

\end{abstract}


\section{INTRODUCTION}\label{introduction}

Gamma-ray bursts (GRBs) are the most luminous explosions in the universe. Their basic
scenario based on the emission from extremely relativistic outflows with bulk Lorentz
factors $~10^{2}-10^{3}$ and isotropic energies of $10^{48}-10^{55} \rm{erg}$ has been
tested, although many questions remain open. The Fermi satellite has advanced our
knowledge of GRBs significantly, while raising some new puzzles. During its first
$\sim2 {\rm yr}$ of operation as of July 27th, 2010, Fermi has observed 19 GRBs 
with photons detected in the LAT (Large Area Telescope) instrument. 
These observations reveal three new properties 
\citep{Granot2010, Abdo080825C, Abdo080916C, Abdo081024B, Abdo090510, Abdo090902B, 
Ackermann090217A, Ackermann090510, Ackermann090926A}:
i) A delayed high energy emission, e.g., 
in GRB 080916C, GRB 081024B, GRB 090510, GRB 090902B and GRB 090926A\footnote{The first
$>100 {\rm MeV}$ photons arrive later than the first lower energy photons detected
by GBM (Gamma-ray Burst Monitor).}.
ii) A temporally extended high energy emission: at least 10 of the first 19 Fermi 
LAT GRBs have long-lived high energy emission, lasting much longer than the burst 
duration in the sub-MeV band (which declines very rapidly);
in 4 out of 10 GRBs, the long-lived LAT light 
curves have a relatively steeper slope, 
for example, $-1.33\pm0.08$ for GRB 080916C, $-1.70\pm0.08$ for GRB 090510,
$ -1.40\pm0.06$ for GRB 090902B, $-2.05\pm0.14$ for GRB 090926A 
according to \citet{Zhang2010catalog}.
iii) A deviation from a pure Band spectral function, showing an extra component in
GRB 090510, GRB 090902B, GRB 090926A.
These are among the brightest bursts, but the observations are compatible with the 
hypothesis of having such a component also in the other, less bright bursts, where 
it is harder to detect {\citep{Granot2010}}.

Among these 19 Fermi LAT GRBs, GRB 090510 is a short, hard burst, with a duration
$T_{90}=0.30\pm0.07\rm s$\citep{DePasquale2010, Ukwatta2009}, located at a redshift 
$z=0.903\pm0.003$ \citep{McBreen2010, Rau2009}. It has been detected by Fermi 
\citep{Guiriec2009, Ackermann090510}, AGILE \citep{Longo2009}, Swift
\citep{Hoversten2009}, Konus-Wind \citep{Golenetskii2009} and Suzaku \citep{Ohmori2009}.
Thus, a large amount of high-quality broadband information is available on this burst,
including optical, X-ray, MeV and GeV emission. The Swift BAT instrument triggered on
GRB 090510 at $T_0^{\rm BAT}=00:23:00.4$ UT, May 10th, 2009 \citep{Hoversten2009},
while the GBM instrument onboard Fermi triggered on at
$T_0^{\rm F}=00:22:59.97$ UT, May 10th, 2009 \citep{Abdo090510}. Thus, there is a
deviation between the two trigger times, which is $\Delta T_0=T_0^{\rm BAT}-T_0^{\rm F}
=0.43 {\rm s}$. 
Hereafter we adopt the BAT trigger time $T_0^{\rm BAT}$ as a natural start time 
$T_0$ for computing the afterglow evolution, this being the onset of the main burst.

The high energy emission of GRB 090510 has all three of the new features we summarized
above: i) the bulk of the photons above 30 MeV arrive $253\pm 34 \rm ms$ later than
those below $1\rm MeV$ \citep{Abdo090510};
ii) the high energy emission above $100\rm MeV$ shows a simple
power law decay lasting $200 {\rm s}$ with a temporal decay index
$\alpha_{\rm LAT}=1.38\pm0.07$ \citep{DePasquale2010}
\footnote{Here we use the convention $F_{\nu}\propto t^{-\alpha}\nu^{-\beta}$.}; 
iii) the time-integrated spectrum from
$T_0+0.07 \rm s$ to $T_0+0.57\rm s$ is best fit by a Band function and a
power-law spectrum \citep{Abdo090510}; the extra power-law component photon index of
$-1.62\pm 0.03$ can fit the data well up to the highest-energy ($31 \rm GeV$) photon 
\citep{Abdo090510}. 

The XRT observations alone give a spectral index $0.57\pm 0.08$ 
\citep{Hoversten2009}, while a detailed analysis of the temporal XRT emission 
combined with the LAT emission indicates a spectral
index $\beta_{\rm X}$ ranging from $0.51$ to $0.81$\citep{DePasquale2010},
and a temporal decay index $\alpha_{{\rm X},1}=0.74\pm0.03$ before a break time
$t_{\rm X,b}=1.43\rm ks$, which
subsequently steepens to $\alpha_{{\rm X},2}=2.18\pm0.10$.
The optical emission initially rises with a temporal index
$\alpha_{{\rm opt},1}=-0.5^{+0.11}_{-0.13}$, and after a break time
$t_{\rm opt,b}=1.58^{+0.46}_{-0.37}\rm{ks}$ it decays with a temporal index
$\alpha_{{\rm opt},2}=1.13^{+0.11}_{-0.10}$ \citep{DePasquale2010}.

Most of the  models aimed at explaining the long-lived high energy
emission of GRB 090510 have favored the view that the high energy
photons arise from the afterglow emission, being generated via
synchrotron emission in the external forward shock
\citep[e.g.,][]{Kumar2009080916C, Kumar20093GRBs, corsi2010, DePasquale2010, Gao2009,
Ghirlanda2010, Ghisellini2010, Wang2010, Razzaque2010}
. This explanation is fairly natural,
since an external forward shock model can account for, at least in
its gross features, not only for the observed delay of the $>100
\rm{MeV}$ photons, which corresponds to the deceleration time-scale
of the relativistic ejecta, but also for the long lasting $>100{\rm
MeV}$ emission, which can be attributed to the power-law decay of
the synchrotron external forward shock emission \citep{Kumar2009080916C, Kumar20093GRBs}.
However, various of the above cited authors use somewhat different
readings of the publicly available spectral and temporal slope data,
which lead them to favor different explanations for the rapid decay
of the long-lived high energy emission, falling into five different
classes of models as follows.

One set of models interprets the LAT emission as synchrotron emission of electrons 
accelerated in a standard adiabatic ISM forward shock 
\citep{Kumar2009080916C, Kumar20093GRBs, corsi2010, DePasquale2010}. 
These authors argue that they can explain the LAT, X-ray, and optical data with 
plausible parameter values (and we revisit these arguments below). 
However, these authors did not {perform} a complete enough
study to confirm whether the whole LAT data including the first $1$ second 
can be explained by this type of models. 
\citet{Kumar2009080916C, Kumar20093GRBs} fit the late LAT data as $F_{\nu }
\propto t^{-1.2}$, but do not give much significance to the early-time 
LAT data. \citet{corsi2010} fit the whole LAT data as $t^{-1.3}$, 
which seems to explain the early-time LAT data but exceeds the $2\sigma$ 
upper limits at late times ($\sim 100-1000s$). \citet{DePasquale2010} 
suggest a steeper decay slope of the high energy emission (evolving 
as $t^{-1.38}$) for the whole LAT data by taking a larger electron distribution 
index $p=2.5$, but they don't take into account in sufficient depth the early-time 
LAT data ($t<1{\rm s}$), which is necessary in order to conclude what is the 
origin of the entire LAT emission (see more discussion in Section 6).

Other models, e.g., \citet{Neamus2010} attribute the high-energy photons to 
synchrotron-self-Compton scattering (SSC) from an adiabatic forward shock 
propagating into a wind-like medium; this, however, requires an extremely 
small magnetic energy fraction $\epsilon_B=10^{-10}$.
Another, different adiabatic forward shock model
analyses in greater detail the Klein-Nishina (KN) effects on the
high energy inverse Compton process \citep{Wang2010}. For some
reasonable parameters, the KN effect, as it weakens in time, results
in the synchrotron high energy emission being increasingly
suppressed by the SSC cooling, which
steepens the synchrotron high energy emission decay slope by a
factor as large as $0.5$. A fourth model views the high-energy
emission as decaying proportional to $t^{-1.5}$, which is
interpreted as being caused by synchrotron emission of electrons
accelerated in a forward shock in the radiative, rather then
adiabatic, regime. For this, the electron population must be
significantly enriched, which is attributed to pair production
between back-scattering photons and prompt outward-going photons
\citep{Ghisellini2010}. This model explains the high energy emission
without considering the constraints from lower energy, e.g., GBM
band, XRT band and UVOT band emission, and the pair formation
becomes inefficient at shock radii larger than $10^{16}{\rm cm}$,
while $R_{\rm dec}\gg10^{16}{\rm cm}$ for this burst (see \S 4). 
A fifth type of model for the afterglow of
this burst (and others) is a hadronic model \citep{Razzaque2010},
which explains the high energy emission as proton synchrotron
emission, while attributing the low energy emission to electron
synchrotron emission from a forward external shock. This requires a
large total kinetic energy $E_{\rm k,iso}=2\times10^{55}\rm
erg$\footnote{Similarly large energies are required for hadronic
models of the prompt emission of this burst \citep{Asano2009}.}, with
a low radiation efficiency and an extremely small fraction of
electron energy $\epsilon_e=10^{-4}$. 

In this article, we re-examine the first set of models (the standard
adiabatic forward shock model) in significantly greater detail than hitherto. 
We present detailed arguments indicating that it is most likely that the forward 
shock synchrotron emission can only explain the LAT emission from $\sim2-3$ sec.
This conclusion disfavors the external shock origin
of the early-time LAT emission \citep{Ghisellini2010, Kumar2009080916C}, 
but supports the suggestions that it is related to the prompt emission 
\citep{corsi2010, DePasquale2010}.
In $\S$ 2, we examine the
XRT and UVOT observations, and set up a model of the long-lived emission based on
synchrotron emission from electrons accelerated in a forward shock in a uniform
ambient environment, including these X-ray and optical/UV observations.
We discuss the impact of the Klein-Nishina effects on the high energy emission,
under the constraints imposed by the lower energy observations, which are found to
be significant for suppressing the SSC cooling. In $\S 3$ we use
a semi-analytical model to calculate the development of the dynamical quantities
of the forward shock across the deceleration time and into the self-similar phase,
and use this to calculate the radiation properties of the long-lived high energy
emission produced by synchrotron emission.
We find a reasonable set of parameters which can explain most of the late afterglow,
except for the six earliest LAT data points in the light curve.  In $\S 4$ we check
several possibilities for the origin of this early high energy emission at times
$t<2-3 \rm s$. In $\S 5$ we discuss the possibility of the line-of-sight prompt emission 
as the origin of the early-time high-energy emission. 
In $\S 6$ we discuss our conclusions concerning the most
probable origin of the high energy emission from the short GRB 090510.

\section{Forward shock model}
\label{sec:FS}

\subsection{Constraints from low energy emission of GRB 090510}
\label{sec:lowen-cons}

The afterglow emission of GRB is generally well explained by synchrotron 
emission from electrons accelerated by the shock produced during a spherical relativistic 
shell colliding with an external medium.
From the spectral index $\beta_{\rm X} \sim 0.51 - 0.81$ and the light 
curve slope $\alpha_{\rm X,1}=0.74\pm 0.03$ \citep{DePasquale2010},  
the closure relation for the X-ray afterglow
$\alpha_{\rm X,1}-1.5\beta_{\rm X}$ ranges from $-0.51$ to $0.01$,
suggesting a slow-cooling ISM external
forward shock model with $\nu_{\rm m}^{\rm f}<\nu_{\rm X}<\nu_{\rm c}^{\rm f}$
\footnote{Hereafter we use the subscripts or superscripts $'\rm f'$ and $'\rm r'$ to
represent the quantities of the forward-shocked and reverse-shocked regions,
respectively, and we use the convention $Q_x=Q/10^x$ in cgs units throughout the paper.},
which implies that the decay index of the X-ray light curve before the jet break is
$\alpha_{\rm X,1}=3(p-1)/4$ and the X-ray spectral index is $\beta_{\rm X}=(p-1)/2$
\citep{Sari1998afterglow}.\footnote{
The closure relation $\alpha_{\rm X,1}-1.5\beta_{\rm X}=-0.5$ indicates the 
slow/fast-cooling ISM/wind external forward shock model with 
$\nu_{\rm X}>\nu_{\rm c}^{\rm f}$ and the spectral index 
$\beta_{\rm X}=p/2$, but this implies $p=1.6<2.0$, which is not favored
by numerical simulations of shock acceleration (e.g., \citet{Achterberg2001})
or by observational data of general GRB afterglows \citep{Freedman2001}.}
From the spectral index we can get a constraint on $p$ which is $p \sim 2.0 - 2.6$. 
In the external shock model, the break seen in X-ray light curve can 
be explained as a jet break. We assume that the jet expands sideways
(Sari et al. 1999). In this case the X-ray light curve slope steepens 
gradually from $-3(p-1)/4$ to $-p$, so that we have $3(p-1)/4 < \alpha_{\rm X,2}
< p$, which is reduced to $2.2 < p < 3.9$ by taking $\alpha_{\rm X,2} \sim 2.2$.
Combining the above two constraints we have $2.2 \lesssim p \lesssim 2.6$. Then
the X-ray decay slope before the break should be $0.9 \lesssim \alpha_{\rm X,1}
\lesssim 1.2$, which can be acceptable if we take into account the observed 
fluctuation of the X-ray flux, as the light curves we will show below.

For the rising portion of the optical light curve before the break time, it is natural
to assume that the optical band is below $\nu_{\rm m}^{\rm f}$, which induces an
optical light curve slope of $t^{1/2}$ 
\citep{Sari1998afterglow, Kumar20093GRBs}.
The predicted spectral slope $\nu^{1/3}$ is consistent with the observations
within the large error bars \citep{DePasquale2010}.
Thus, one can try to explain both the X-ray emission and the optical emission before
the break time with synchrotron emission of electrons accelerated in an adiabatic
external forward shock with the assumptions $\nu_{\rm opt}<\nu_{\rm m}^{\rm f}$ and
$\nu_{\rm m}^{\rm f}<\nu_{\rm X}<\nu_{\rm c}^{\rm f}$.
Then, assuming that the optical band is in the regime of 
$\nu_{\rm m}^{\rm f}<\nu_{\rm opt}< \nu_{\rm c}^{\rm f}$ after the break time, 
the slope of the optical 
post-break light curve is the same as that of the X-ray light curve. 

In the analytical calculations of this section, we assume that the self-similar phase
conditions have been established, and for simplicity we neglect the structure of the
shock wave, considering a spherical shock with a total isotropic energy $E$ and a Lorentz factor
$\Gamma$.  At late times, the adiabatic dynamical evolution of the spherical shock is
in the Blandford $\&$ Mckee self-similar phase, where $E=16\pi\Gamma^2R^3nm_pc^2/17$ is
constant and the scaling law of the shock wave is $\Gamma\propto R^{-3/2}$
\citep{Blandford1976}. The shock propagates a distance 
$\delta R\sim 2\Gamma^2c\delta t/(1+z)$ during the small observing time 
$\delta t$ \citep{Sari1997dyn}, and integrating this and using the scaling
law, one obtains $t=(1+z)R/8\Gamma^2 c$.

According to \citet{Sari1998afterglow}, the cooling and minimum Lorentz factors of the 
electrons in the forward shock depend
on the total isotropic kinetic energy $E$, the number density of the external environment $n$,
and the fraction of the electron energy and magnetic field energy $\epsilon_{e\rm f}$
and $\epsilon_{B\rm f}$, which can be expressed as
\begin{equation}\label{gammac}
\gamma_{\rm c}^{\rm f}=\frac{6\pi(1+z)m_ec}{\sigma_{\rm T}B'^2\Gamma t}=2.0\times10^8\epsilon_{B\rm 
f,-4}^{-1}E_{53.5}^{-3/8}n_{-4}^{-5/8}t_2^{1/8}\left[{1+Y(\gamma_{\rm c}^{\rm f})}\right]^{-1}
\end{equation}
and
\begin{equation}\label{gammam}
\gamma_{\rm m}^{\rm f}=\epsilon_{e\rm f}\frac{p-2}{p-1}\frac{m_p}{m_e}\Gamma=9.0\times 10^4
\epsilon_{e\rm f,-0.4}E_{53.5}^{1/8}n_{-4}^{-1/8}t_2^{-3/8}g_p,
\end{equation}
respectively, where 
\begin{equation}\label{B}
B'=(32\pi\epsilon_{B\rm f}\Gamma^2nm_pc^2)^{1/2}=
1.4\times10^{-2}\epsilon_{B\rm f,-4}^{1/2}E_{53.5}^{1/8}n_{-4}^{3/8}t_{2}^{-3/8}{\rm G}
\end{equation} 
and  $g_p\equiv 3(p-2)/(p-1)$, $p$ is the power-law index of the electron 
energy distribution, $Y(\gamma_{\rm c}^{\rm f})$ is the Compton 
parameter of the electrons with Lorentz factor $\gamma_{\rm c}^{\rm f}$, 
and $\sigma_{\rm T}$ is Thomson cross section. The cooling and minimum frequencies of electrons are
\begin{equation}\label{hnuc}
h\nu_{\rm c}^{\rm f}=h\frac{q_eB'}{2\pi m_ec}(\gamma_{\rm c}^{\rm f})^2\Gamma=2.3\times 10^{9}\left
[{1+Y(\gamma_{\rm c}^{\rm f})}\right]^{-2}\epsilon_{B\rm f,-4}^{-3/2}E_{53.5}^{-1/2}n_{-4}^{-1}t_2^{-1/2}
{\rm eV}
\end{equation}
and
\begin{equation}\label{hnumin}
h\nu_{\rm m}^{\rm f}=h\frac{q_eB'}{2\pi m_ec}(\gamma_{\rm m}^{\rm f})^2\Gamma=5.0\times 10^
{2}\epsilon_{e\rm f,-0.4}^{2}
\epsilon_{B\rm f,-4}^{1/2}E_{53.5}^{1/2}t_2^{-3/2}g_p^2 {\rm eV},
\end{equation}
respectively. The peak flux density of the forward shock synchrotron emission is
\begin{equation}\label{Fmax}
F_{\rm max}^{\rm f}=\frac{(1+z)N_e^{\rm f}m_e c^2\sigma_{\rm T}B'\Gamma}{12\pi q_e d_{\rm L}^2}=137 
\epsilon_{B\rm f,-4}^{1/2}E_{53.5}n_{-4}^{1/2}\rm{\mu Jy},
\end{equation}
where the total number of the electrons that the forward shock swept up is $N_e^{\rm f}=\frac{16}{17}\pi 
R^3n=1.5\times10^{51}E_{53.5}^{3/4}n_{-4}^{1/4}t_2^{3/4}$, and $d_{\rm L}$ is the luminosity distance.

We get two constraints from the UVOT and XRT data as follows:\\
(i) The optical flux density is about $20 {\rm \mu Jy}$ at $t\sim 100 {\rm s}$
\citep{DePasquale2010}, which indicates that
\begin{equation}\label{fopt}
F_{\rm opt}(t=100\rm s)\simeq F_{\rm max}^{\rm f}\left(\frac{4.5{\rm eV}}{h\nu_{\rm m}^{\rm f}}\right)^{1/3}
=28.5\epsilon_{ef,-0.4}^{-2/3}\epsilon_{Bf,-4}^{1/3}E_{53.5}^{5/6}n_{-4}^{1/2}g_{p}^{-2/3}{\rm \mu Jy}\sim20
{\rm \mu Jy}.
\end{equation}
(ii) The X-ray flux density is about $20 {\rm \mu Jy}$ at $t\sim100 {\rm s}$, which indicates
\begin{equation}\label{fxrt}
F_{\rm XRT}(t=100{\rm s})\simeq F_{\rm max}^{\rm f}\left(\frac{3000{\rm eV}}{\nu_m^{\rm f}}\right)^{-(p-1)/2}
=35.7\epsilon_{ef,-0.4}^{p-1}\epsilon_{Bf,-4}^{(p+1)/4}E_{53.5}^{(p+3)/4}n_{-4}^{1/2}f_{e1}^{3.1}g_P^{p-1}{\rm 
\mu Jy}\sim20{\rm \mu Jy}.
\end{equation}
Combining the above two equations (\ref{fopt}) and (\ref{fxrt}) , we can express the 
fraction of the magnetic field energy and the number density as
\begin{equation}\label{eBcon}
\epsilon_{B\rm f,-4}\sim 0.66\epsilon_{e\rm f,-0.4}^{-4} E_{53.5}^{-1}g_p^{-4}g_{e2}^{12} ,
\end{equation}
\begin{equation}\label{ncon}
n_{-4}\sim 0.66\epsilon_{e\rm f,-0.4}^{4} E_{53.5}^{-1}g_p^4g_{e2}^{-8.0}
\end{equation}
where $g_{e1}=e^{p-2.5}$ and $g_{e2}=e^{(p-2.5)/(3p-1)}$
and we can naturally get $\epsilon_{e\rm f}>0.04E_{53.5}^{-1/4}g_{e2}^3g_p^{-1}$ from
the condition $\epsilon_{B\rm f}<1$.

Inserting equations (\ref{eBcon}) and (\ref{ncon}) into equation (\ref{B}), 
the downstream magnetic field strength is thus constrained to be
\begin{equation}
B'\sim 10\epsilon_{e\rm f,-0.4}^{-1/2}E_{53.5}^{-3/4}t_{2}^{-3/8}g_{p}^{-1/2}g_{e2}^{3.0}{\rm mG}.
\end{equation}
If the shocks involve a magnetic field amplification factor $f_B\geq 1$ (in addition to
shock compression), the upstream magnetic field strength would be
\begin{equation}
B_{\rm u}=B'/(4\Gamma f_{B})\sim 7 f_{B}^{-1} E_{53.5}^{-1}g_{e2}^{2.0}{\rm \mu G}.
\end{equation}
Thus, the upstream magnetic field strength could be $\lesssim 5-10 \mu$G
\citep[e.g.,][]{Kumar20093GRBs}, apparently compatible with shock-compression of the
typical magnetic field in the interstellar medium, with no need of magnetic field
amplification. However, as we discuss in \S \ref{sec:disc}, the external density
deduced here is much below the average interstellar value, and likely so would be
the external magnetic field, so that additional field amplification may be needed.

Inserting equations (\ref{eBcon}) and (\ref{ncon}) into equations (\ref{hnuc}), 
(\ref{hnumin}) and (\ref{Fmax}), the characteristic energies and the peak flux 
density of synchrotron emission are therefore
\begin{equation}
h\nu_{\rm c}^{\rm f}\sim 6.6\times 10^{9}\epsilon_{e\rm f,-0.4}^{2} E_{53.5}^2t_2^{-1/2}[1+Y(\gamma_{\rm c}
^{\rm f})]^{-2}g_p^2g_{e2}^{-10}{\rm eV},
\end{equation}
\begin{equation}\label{numin2}
h\nu_{\rm m}^{\rm f}\sim4.0\times 10^2 t_2^{-3/2} g_{e2}^{6.0}{\rm eV},
\end{equation}
and
\begin{equation}
F_{\rm max}^{\rm f}\sim90 g_{e2}^{2.0} {\rm \mu Jy},
\end{equation}
respectively. The above equations show that
$\nu_{\rm opt}<\nu_{\rm m}^{\rm f}$ until $t\sim 1.4{\rm ks}$, and
$\nu_{\rm m}^{\rm f}<\nu_{\rm X}<\nu_{\rm c}$ under the constraint 
$\epsilon_{e\rm f,-0.4}E_{53.5}>1.2\times10^{-3}t_{2}^{1/4}g_p^{-1}g_{e2}^{5.0}$, 
which is easy to satisfy, where $Y(\gamma_{\rm c}^{\rm f})\ll1$ as discussed in 
\S \ref{sec;kn}.  Thus, they are consistent with our previous assumptions.
 The radiative efficiency in the assumed slow cooling regime is \citep{Sari2001}
 \begin{equation}\label{epsilon}
 \eta_r=\epsilon_{e\rm f}\left(\frac{\nu_{\rm m}^{\rm f}}
{\nu_{\rm c}^{\rm f}}\right)^{\frac{p-2}{2}}=6.3\times10^{-3}\epsilon_{e\rm f,-0.4}^{3-p}
E_{53.5}^{2-p}t_2^{-(p-2)/2}g_{p}^{2-p}g_{e1}^{-8.3}g_{e2}^{8.0(p-2)},
\end{equation}
which is much less than unity, consistent with our previous assumption of an adiabatic
forward shock model.

In order to check whether the LAT emission predicted by synchrotron emission from
an adiabatic forward shock can explain the LAT observations we need to study two 
situations. 

(i)
Under the condition $\epsilon_{e\rm f,-0.4}E_{53.5}>0.39g_p^{-1}g_{e2}^{5.0}$, 
we yield $\nu_{\rm m}^{\rm f}<1{\rm GeV}<\nu_{\rm c}^{\rm f}$ at $t\la100{\rm s}$, 
the average synchrotron flux density from the forward shock
in the LAT band ($100{\rm MeV}$ to $4 {\rm GeV}$) is
\begin{equation}\label{Flat}
F_{\rm LAT}^{\rm f}\sim F_{\rm max}^{\rm f} \left(\frac{1{\rm GeV}}{h\nu_{\rm m}^{\rm f}}\right)^{-\frac{p-1}
{2}}\sim 1.4\times 10^{-3} t_2^{-3(p-1)/4} g_{e1}^{-6.4} {\rm \mu Jy}.
\end{equation}
which is independent of the electron energy $\epsilon_{e\rm f}E$.  In addition, 
the slope of the LAT light curve is also constrained, as $\alpha_{\rm LAT}=3(p-1)/4$, 
the same as that of the X-ray light curve.  For $p=2.5$, $\nu_{\rm m}^{\rm f}<
1{\rm GeV}<\nu_{\rm c}^{\rm f}$ at $t\la100{\rm s}$ is satisfied if 
$\epsilon_{e\rm f,-0.4}E_{53.5}>0.39$, then the predicted LAT flux is 
$1.4\times10^{-3}{\rm \mu Jy}$ at $100$ s, consistent with the 
observational data within the error bars at around $100$ s. 
The corresponding slope of the predicted LAT light curve is around $-1.125$, which 
can explain the late-time data of the LAT observation.
If we take a small electron index, for example, $p\sim 2.2$, we have 
$\nu_{\rm m}^{\rm f}<1{\rm GeV}<\nu_{\rm c}^{\rm f}$ at $t=100{\rm s}$ if 
$\epsilon_{e \rm f,-0.4}E_{53.5}>0.60$. Then the predicted LAT flux can be calculated 
by equation (\ref{Flat}), approximated as $\sim 8.9\times 10^{-3}{\mu \rm Jy}$ at $100$
s, which is almost one order of magnitude smaller than the observed LAT flux.
What's more, the slope of the predicted LAT light curve is about $-0.9$, which is 
too shallow to explain the late-time LAT observation.  Thus, in the case 
$\nu_{\rm m}^{\rm f}<1{\rm GeV}<\nu_{\rm c}^{\rm f}$ at $100$ s, we can
exclude the $p=2.2$ model.

(ii) If the electron energy satisfies the condition $\epsilon_{e\rm f,-0.4}E_{53.5}
\leq0.12g_p^{-1}g_{e2}^{5}$, the lower end of the LAT band, $100$ MeV, 
is above the frequency $\nu_c^{\rm f}$, 
and the predicted LAT flux at $t=100{\rm s}$ can be calculated as
\begin{equation}\label{Flatsteeper}
F_{\rm LAT}^{\rm f}\sim F_{\rm max}^{\rm f}
\left(\frac{h\nu_{\rm c}^{\rm f}}{h\nu_{\rm m}^{\rm f}}\right)^{-\frac{p-1}{2}}
\left(\frac{1{\rm GeV}}{h\nu_{\rm c}^{\rm f}}\right)^{-\frac{p}{2}}\sim
3.6\times10^{-3}\epsilon_{ef,-0.4}E_{53.5}t_{2}^{-(3p-2)/4}g_pg_{e1}^{-7.4}
g_{e2}^{3p-6}{\rm \mu Jy}.
\end{equation}
Inserting the condition $\epsilon_{e\rm f,-0.4}E_{53.5}\leq0.12g_p^{-1}g_{e2}^{5}$ into 
equation (\ref{Flatsteeper}), we see that $F_{\rm LAT}^{\rm f}\leq4.3\times 10^{-4}
g_{e1}^{-8.4}t_2^{-(3p-2)/4} {\rm \mu Jy}$. For $p=2.5$, the predicted LAT flux at 
$100$ s is $F_{\rm LAT}^{\rm f}(100{\rm s})\leq4.3\times 10^{-4}{\rm \mu Jy}$, which 
is almost one order of magnitude lower than the observed LAT flux.  Thus in the 
case $\nu_{\rm LAT}>\nu_{\rm c}^{\rm f}$ at $100$ s, we can exclude the $p=2.5$ 
model.

For $p=2.2$, to be consistent with the observed LAT flux at $100$ s, 
the condition $\epsilon_{ef,-0.4}E_{53.5}=6.3\times10^{-2}$ is required, 
constraining the electron energy to a very small value. Here we adopt fairly standard 
values of $\epsilon_{e\rm f}=0.16 - 0.6$, similar to those observed in long GRBs, 
since these are determined by collisionless shock physics processes on microphysical 
scales, which should be independent of the global properties of GRBs.
Then the constrained total kinetic energy could be $(0.13-0.47)\times 10^{53}{\rm erg}$.
Since the isotropic energy at $10{\rm keV}-30{\rm GeV}$ energy band during 
$T_0+0.03{\rm s}-T_0+0.53{\rm s}$ is $E_{\rm iso}=(1.08\pm0.06)\times 10^{53}{\rm erg}$ 
\citep{Ackermann090510}, the radiative efficiency of the prompt emission 
is in the range $70\%-89\%$.  
Moreover, we constrain $\epsilon_{Bf,-4}=88\epsilon_{e,-0.4}^{-3}$ and 
$n_{-4}=1.0\epsilon_{e,-0.4}^{5}$ by inserting the constraint 
$\epsilon_{ef,-0.4}E_{53.5}=6.3\times10^{-2}$ into equations 
(\ref{eBcon}) and (\ref{ncon}).
The constrained $\nu_{\rm c}^{\rm f}$ is $h\nu_{\rm c}^{\rm f}=1.1\times10^7{\rm eV}$ at $100{\rm s}$, 
and $\nu_{\rm c}^{\rm f}<10^8{\rm eV}$ when $t>1.2{\rm s}$.
Since the expected slope of the LAT light curve is $\alpha_{\rm LAT}=(3p-2)/4=1.15$ 
for $p=2.2$ in the $\nu_{LAT}>\nu_{\rm c}^{\rm f}$ case, similar to that expected from
the $p=2.5$ model in the $\nu_{\rm m}^{\rm f}<\nu_{\rm LAT}<\nu_{\rm c}^{\rm f}$ case, 
this can also explain the slope of the late-time LAT light curve.
Even though $\nu_{\rm m}^{\rm f}$ decreases by a factor of $1.5$ by taking $p=2.2$ 
rather than taking $p=2.5$, it doesn't change the optical break time by much,
and we may not rule out the $p=2.2$ model in the $\nu_{\rm LAT}>\nu_{\rm c}^{\rm f}$ 
case.

Thus, both the $p=2.5$ model in the $\nu_{\rm m}^{\rm f}<1{\rm GeV}<\nu_{\rm c}^{\rm f}$ 
case and the $p=2.2$ model in the $\nu_{\rm LAT}>\nu_{\rm c}^{\rm f}$ case
can explain the late-time LAT light curve, leading to the 
same conclusion, since the predicted LAT light curves have the similar slope. For 
presentation purposes, hereafter, we discuss the $p=2.5$ model in the 
$\nu_{\rm m}^{\rm f}<1{\rm GeV}<\nu_{\rm c}^{\rm f}$ case.

\subsection{Impact of Klein-Nishina effects on the constraints}
\label{sec;kn}

\citet{Wang2010} have studied the Klein-Nishina (KN) effects on
high-energy gamma-ray emission in the early afterglow, and find that
at early times the KN suppression on the IC scattering cross section
for the electrons that produce the high-energy emission is usually
strong, and therefore their inverse-Compton losses are small, with a
Compton parameter $Y$ of less than a few for a wide range of
parameter space. This leads to a relatively bright synchrotron
afterglow emission at high energies at early times. However, as the
KN effects weaken with time, the inverse-Compton losses increase and
the synchrotron high energy emission is increasingly suppressed,
which leads to a more rapid decaying synchrotron emission. This
provides a potential mechanism for the steep decay of the
high-energy gamma-ray emission seen in some Fermi LAT GRBs.

The Compton  parameter for electrons with Lorentz factor $\gamma_e$ is defined as the
ratio of the synchrotron self-inverse Compton (SSC) to the synchrotron emissivity, i.e.
\begin{equation}
Y(\gamma_e)\equiv\frac{P_{\rm SSC}(\gamma_e)}{P_{\rm syn}(\gamma_e)}.
\end{equation}
When the KN suppression on the scattering cross section is
negligible, $Y({\gamma_e})= Y(\gamma_c)$ is a constant for the
slow-cooling case \citep{Sari2001}. However, for high energy
electrons with a significant KN effect, $Y(\gamma_e)$ is no
longer a constant and this affects the electron radiative cooling
function, as well as the continuity equation of the electron
distribution. The self-consistent electron distribution is given by
\begin{equation}\label{n_slow}
N(\gamma_e)=\left\{{}
\begin{array}{ll}
C_1\gamma_e^{-p} \,\,\,\,\,\,\,\,\,\,\,\,\,\,\,\,\,\,\,\,\,\,\,\,\,\,\,\,\,\,\,\,\,\,\,\, \gamma_m<
\gamma_e<\gamma_c \\
\frac{1+Y(\gamma_c)}{1+Y(\gamma_e)}C_1\gamma_c \gamma_e^{-p-1}
\,\,\,\,\,\, \gamma_c<\gamma_e
\end{array}\right .
\end{equation}
for the slow-cooling  case \citep{Nakar2009, Wang2010},
where $C_1$ is a constant. The high energy synchrotron
photons with energy $h\nu_*$ are produced by electrons with Lorentz
factor $\gamma_*$ which typically have $\gamma_*>max(\gamma_{\rm c},
\gamma_{\rm m})$. Thus, the number density of electrons of
$\gamma_*$  is
\begin{equation}
N(\gamma_*)=\frac{1+Y(\gamma_c)}{1+Y(\gamma_*)}C_1\gamma_c
\gamma_*^{-p-1}=\frac{N_{\rm syn}(\gamma_*)}{1+Y(\gamma_*)},
\end{equation}
where $N_{\rm syn}(\gamma_*)=C_1\gamma_{c}[1+Y(\gamma_c)]\gamma_*^{-p-1}$
is the number density of electrons of $\gamma_*$ when only the synchrotron cooling
is considered \citep{Sari1998afterglow}. Therefore, the number density of electrons
with Lorentz factor $\gamma_*$ is a factor of $1+Y(\gamma_*)$ lower than that in the
case where only the synchrotron cooling is considered. Thus the synchrotron luminosity
is correspondingly reduced by the same factor. We have $Y(\gamma_*)\propto t^{1/2}$
, in the slow cooling regime, as long as $\gamma_*>\Gamma m_{e}
c^2/h\nu_{\rm m}^{\rm f}$ \citep{Wang2010}. In that case, if
$Y(\gamma_*)\gg 1$ the synchrotron luminosity is suppressed by the
factor $Y(\gamma_*)$ which is in proportion to $t^{1/2}$, i.e. 
the light curve decay of the high energy synchrotron emission
could be steepened by a factor $1/2$ at most.  Meanwhile, the
distribution of electrons which are in the region
$\gamma_{\rm m}<\gamma<\gamma_{\rm c}$ is not affected by $Y(\gamma)$, due to
equation \ref{n_slow}, thus the lower energy synchrotron emission
decay slope is normal.

The critical photon energy above which the scattering with electrons of energy $\gamma_e$
just enters the KN scattering regime is defined as $h\nu_{\rm KN}(\gamma_e) \equiv
\Gamma m_{e}c^2/\gamma_e$, i.e. 
\begin{equation}
h\nu_{\rm KN}(\gamma_{\rm c}^{\rm f})=\Gamma m_{e}c^2/\gamma_{\rm c}^{\rm f}=0.52\epsilon_{e\rm 
f,-0.4}^{-2}E_{53.5}^{-1}t_2^{-1/2}g_p^{-2}g_{e2}^{8.0}[1+Y(\gamma_{\rm c}^{\rm f})]{\rm eV}.
\end{equation}
which is much smaller than $\nu_{\rm m}^{\rm f}$ under the condition 
$\epsilon_{e\rm f,-0.4}^{2}E_{53.5}>0.13t_4g_p^{-2}g_{e2}^{2.0}
[1+Y(\gamma_{\rm c}^{\rm f})]$.
In this case the synchrotron-self Compton scattering is strongly suppressed due to
KN effects, and
\begin{eqnarray}
Y(\gamma_{\rm c}^{\rm f})\left[1+Y(\gamma_{\rm c}^{\rm f})\right]
&=&\frac{\epsilon_{e\rm f}}{\epsilon_{B\rm f}}\left(\frac{\gamma_{\rm c}^{\rm f}}{\gamma_{\rm m}^{\rm 
f}}\right)^{2-p}\left(\frac{\nu_{\rm m}^{\rm f}}{\nu_{\rm c}^{\rm f}}\right)^{(3-p)/2}\left(\frac{\nu_{\rm KN}
(\gamma_{\rm c}^{\rm f})}{\nu_{\rm m}^{\rm f}}\right)^{4/3} \nonumber \\
&=&1.4\times 10^{-4}\left[1+Y(\gamma_{\rm c}^{\rm f})\right]^{7/3}\epsilon_{e\rm f,-0.4}^{4/3}E_{53.5}^{-4/3}
t_{2}^{5/6}g_{p}^{1/3}g_{e2}^{-1.3}.
\end{eqnarray}
From the above equation, we find that $Y(\gamma_{\rm c}^{\rm f})\ll1$ under the condition 
$\epsilon_{e\rm f,-0.4}^{-1}E_{53.5}\gg 0.023t_4^{5/8}g_p^{1/4}g_{e2}^{-0.98}$, 
which is easy to be satisfied before $t<10^{4}{\rm s}$.
Therefore, since $Y(\gamma_{*})\leqslant Y(\gamma_c)\ll1$,
the distribution of electrons which contribute to the high-energy synchrotron photons doesn't change,
according to equation (\ref{n_slow}), which means that the decay slope
of synchrotron high energy emission can not be affected by
$Y(\gamma_{*})$. 

\section{Forward Shock Synchrotron Emission Evolution}
\label{sec:FSsyn}

We adopt the following parameters for the calculations.
From equation (\ref{Flat}), we see that the flux density in the LAT range is independent of the total 
isotropic kinetic energy and the electron energy fraction at $t<100{\rm s}$, under the constraint
$\epsilon_{e\rm f,-0.4}E_{53.5}>0.39$. Thus, we can fix the  total energy of the afterglow 
as $E=3.0\times10^{53} {\rm erg}$, indicating a radiation efficiency as $25\%$.
Even if we choose other values of the total kinetic energy 
under the constraint $\epsilon_{e\rm f,-0.4}E_{53.5}>0.39$, 
our conclusion will not change.
Then the fraction of electron energy is constrained as $\epsilon_{e\rm f}>0.16$.
We adopt fairly standard values of $\epsilon_{e\rm f}\geq0.16$.
From the equations (\ref{eBcon}) and (\ref{ncon}), we can get the values of
$\epsilon_{B\rm f}=2.6\times 10^{-3} - 1.3\times10^{-5}$, corresponding to an external 
density range
$\left[ n=1.7\times10^{-6} {\rm cm}^{-3} - 3.3\times10^{-4}{\rm cm^{-3}}\right]$.
These densities range from much less than the intergalactic medium (IGM), up to
a low density inter-cluster medium (ICM) or possibly galactic halo baryon density.
The critical Lorentz factor which is the boundary between the thin and thick shell cases
\citep{Kobayashi2007} is $\Gamma_c=[3(1+z)^3 E/32\pi nm_pc^5T^3]^{1/8}\sim 5.2 \times
10^3 E_{53.5}^{1/8}n_{-4}^{-1/8}T_{-0.5}^{-3/8}$, where the duration of GRB 090510   is
$T=0.30\pm0.07{\rm s}$. For an initial Lorentz factor which is not too large,
it is reasonable to assume that the initial Lorentz factor is smaller than the
critical Lorentz factor, which means the shell has given the ambient medium an 
energy comparable to its initial energy at the deceleration time
$t_{\rm dec}=\left[{3(1+z)^3E}/{32\pi c^5nm_p\Gamma_0}\right]^{1/3}>T$,
indicating the thin shell case.
Then the initial Lorentz factor of the forward shock can be expressed as a function of
the deceleration time, which is $\Gamma_0=3.3\times 10^3 E_{53.5}^{1/8}n_{-4}^{-1/8}
t_{\rm dec}^{-3/8}$.
If, for example, we take an initial Lorentz factor as suggested
by previous Fermi analyses of $2.2\times 10^3 - 5.2\times 10^3$, and a not too small
number density of $n \sim 10^{-4}{\rm cm^{-3}}$, we would obtain a deceleration time
of $t_{\rm dec}\sim 0.3 {\rm s} - 3 {\rm s}$.

To calculate the dynamics of the evolution of the blast wave including the transition
from the quasi-free expansion, through the deceleration and into the self-similar
phase, we use the relativistic hydrodynamics equations for the evolution of the shock
radius $R$, the mass $m$ swept up by the shock, the opening half-angle of jet $\theta$
and the Lorentz factor of the shock $\Gamma$ \citep[e.g.,][]{Huang2000apj} (see
Appendix A), and solve these equations numerically.
The solution of these equations provides the dynamical quantities which we use to
calculate the radiation spectrum and the light curves discussed in the following.
We take into account also the jet evolution as it goes through the jet break and
starts to expand sideways. The half-angle of the jet evolves as
$d\theta_{\rm j}=({c_{\rm s}dt'}/{r})=[{(1+\beta)\Gamma c_{\rm s}}/{r}]dt$, with a spreading
velocity in the comoving frame approximated by the sound speed $c_{\rm s}$
\citep{Rhoads1997, Rhoads1999, Huang2000mn}.  
After the inverse Lorentz factor
becomes  larger than the initial jet angle, the spreading of the jet speeds up the
shock deceleration significantly, which leads to a steeper light curve.

In figures 1 and  2 we present model light curves in the LAT, XRT and UVOT bands for
two  choices of parameters. The forward shock light curves (solid black lines) are
shown, for each density choice, for two different Lorentz factors and jet opening
angles.  A smaller opening angle is chosen for a larger Lorentz factor in order to 
get a certain jet break time.

In figure 1, we show a solution with a density of $n=1\times10^{-6} {\rm cm}^{-3}$,
which would correspond to a sub-average density IGM environment.
Under the assumptions $\nu_{\rm opt}<\nu_{\rm m}^{\rm f}$ before the optical break
time and $\nu_{\rm m}^{\rm f}<\nu_{\rm X}<\nu_{\rm c}^{\rm f}$, the steep decay of
the latter parts of the X-ray and optical light curves can be explained using such a
forward shock synchrotron emission going through a jet spreading phase, 
as shown in figure 1 and 2. Note, however, that, as in other analyses too,
the late-time optical data are challenging to fit.
This leads to a model forward shock light curve and spectral fit to the X-ray and
optical observations satisfying all the constraints of \S \ref{sec:lowen-cons}, for a set
of parameters $\epsilon_{e\rm f}=0.17$, $E=3\times 10^{53}{\rm erg}$, $p=2.5$,
$\epsilon_{B\rm f}=7.0\times 10^{-3}$, and $\theta_0=0.45^{\circ}$ for $\Gamma_0=4500$,
or $\theta_0=0.38^{\circ}$ for $\Gamma_0=9000$.
The collimation corrected energies are $4.6\times10^{48}$ and $3.3\times10^{48}$ erg.

In figure 2 we show the forward shock solutions for a more moderate density of
$n=10^{-3}{\rm cm}^{-3}$, corresponding to a galactic halo or interarm medium,
which satisfy the low energy constraints as well as the afterglow epoch LAT, XRT
and UVOT data points. The parameters are 
$\epsilon_{e\rm f}=0.6$, $\epsilon_{B\rm f}=1\times10^{-5}$,
$p=2.5$ and $E=3\times 10^{53}$ erg, for a choice of
$\theta_{0}=1.3^{\circ}$ and $\Gamma_0=1900$, and also for a choice of
$\theta_{0}=1.1^{\circ}$ and $\Gamma_0=3800$.
The collimation corrected energies are $3.9\times10^{49}$ and $2.8\times10^{49}$ erg,
which is a typical energy for a short GRB. Despite the larger $\epsilon_{e\rm f}$ in this
fit, the forward shock is still in the adiabatic regime according to equation 
(\ref{epsilon}).

In Figures 1 and 2 the LAT energy band light curves predicted by the forward shock
synchrotron model shown in black solid lines (thicker for the larger Lorentz factor
and smaller opening angle choice, thinner for the smaller Lorentz factor and larger 
angle choice).  These appear to explain well the late time LAT emission, i.e. after 
3 seconds, either with the larger or smaller angle/Lorentz factor, 
no matter how small the deceleration time is, but they fall below the first six LAT data points.
This difference between the early time observed LAT data and the synchrotron forward
shock emission suggests that there may be another radiation component (whose contribution
can be represented by the gray dotted lines in Figures 1 and 2, corresponding to the 
high latitude emission of the prompt emission with variability timescale $\Delta t=0.5{\rm s}$), 
in addition to the forward synchrotron contribution (shown as the black solid lines),
at least for its decaying portion.
The gray solid lines are the sum of the steep decay component and the forward synchrotron
emission with a larger Lorentz factor (the first, lower data point could be part of
either a rising portion or a variable portion of the gray dotted component).
Thus, it is possible that the early-time high-energy emission is not from the
afterglow forward shock synchrotron emission, instead having a different origin.
In the next section we discuss possible origins for the first six data points in
the LAT band around the deceleration time.

\section{Other Possible Components around the Deceleration Time}
\label{sec:tdec}

Besides the forward shock synchrotron emission, around the deceleration time
there are several other possible emission components which could contribute
to the early LAT observations. Their importance can be estimated using approximate values
for the characteristic quantities at the deceleration time. 
At the deceleration time, the Blandford-McKee self-similar solution is 
not yet applicable.
For this we can take
the usual value of the energy as $E=\frac{4 \pi}{3}R^3\Gamma^2nm_pc^2$ and
$R=2\Gamma^2 ct/(1+z)$ as the radius at the deceleration time 
\citep{Sari1998afterglow}, 
from which we obtain 
the initial Lorentz factor of the shock and the deceleration radius,
which are
\begin{equation}
\Gamma_{\rm 0}=3.3\times10^3 E_{53.5}^{1/8} n_{-4}^{-1/8} t_{\rm dec}^{-3/8}
\end{equation}
\begin{equation}
R_{\rm dec}=3.5\times10^{17}E_{53.5}^{1/4}n_{-4}^{-1/4}t_{\rm dec}^{1/4}{\rm cm}.
\end{equation}

Before we examine the other possible components around the deceleration time, we confirm that 
the forward shock synchrotron emission cannot explain the observation at early times in more details.
We can then estimate the approximate high energy emission at the deceleration time
under the previous constraints provided by the  low energy observations. Here we take
the Lorentz factor and radius at the deceleration time as $\Gamma_{\rm d}\sim\Gamma_{\rm 0}/\sqrt{2}$
and $R_{\rm d}\sim R_{\rm dec}/\sqrt{2}$, which
are close to the values indicated by the numerical evolution of the previous section.
Inserting the above $\Gamma_{\rm d}$ and $R_{\rm d}$ into the first expressions of
equations (\ref{gammac}) $-$ (\ref{Fmax}), 
and meanwhile adopting equations (\ref{eBcon}) and (\ref{ncon}), 
the characteristic frequencies and peak flux density of the forward shock synchrotron
emission at the deceleration time are then
\begin{equation} 
h\nu_{\rm m,d}^{\rm f}\sim6.8\times 10^5 t_{\rm dec}^{-3/2}g_{e2}^{6.0}{\rm eV},
\end{equation}
\begin{equation}
h\nu_{\rm c,d}^{\rm f}\sim 3.9\times 10^{10}\epsilon_{e\rm f,-0.4}^{2} E_{53.5}^2t_{\rm dec}^{-1/2}g_p^2 g_
{e2}^{-10}{\rm eV},
\end{equation}
\begin{equation}
F_{\rm max,d}^{\rm f}\sim12g_{e2}^{2.0} {\rm \mu Jy}.
\end{equation}

\citet{Ghisellini2010} suggests the radiative forward shock model to explain the
steep temporal decay of high energy emission, which requires $\nu_{\rm m}^{\rm f}>
\nu_{\rm c}^{\rm f}$ to get the fast cooling case. However, the  radiative model cannot
explain the shallower decay at later time ($t>10{\rm s}$) (as seen in Figure 7 in
\citet{Ghisellini2010}), which agrees better with an adiabatic model in slow cooling
case at that time. According to the constrained characteristic frequencies, the fast
cooling case at the deceleration time requires that
\begin{equation}
\epsilon_{e\rm f}<1.7\times10^{-3}E_{53.5}^{-1}t_{\rm dec}^{-1/2}g_{p}^{-1}g_{e2}^{8.0},
\end{equation}
which is inconsistent with their assumption of a very high energy fraction of electrons
$\epsilon_{e\rm f}=0.9$.  This is because \citet{Ghisellini2010} assume the late-time
emission to be also in the fast cooling case, and do not take into account the
constraints from the low energy emission.

In addition, using equations (\ref{eBcon}) and (\ref{ncon}),
we get an estimate for the forward shock synchrotron emission at $1{\rm GeV}$ at the deceleration time,
\begin{equation}\label{F_f}
F_{1{\rm GeV,d}}^{\rm f}=F_{\rm max,d}^{\rm f}\left(\frac{1{\rm GeV}}{h\nu_{\rm m,d}^{\rm f}}\right)^{-
\frac{p-1}{2}}\sim  0.049 t_{\rm dec}^{-3(p-1)/4} g_{e1}^{-2.6} {\rm \mu Jy}.
\end{equation}
This is almost one order of magnititude below the observed LAT flux density,
for the case when the deceleration time is smaller than $3 {\rm s}$
(this is seen also in Figures 1 and 2).
The synchrotron forward shock emission cannot explain the early-time
high-energy emission (the first six LAT data points near the deceleration time),
although it can explain very well the late LAT results (beyond the sixth LAT data
point, when the afterglow can be considered as established), which are consistent
with the numerical results in \S \ref{sec:FSsyn}.

Thus, we find that there has to be another component, 
contributing to the early high energy emission.
In the rest of this section we consider several such possibilities,
such as a reverse shock synchrotron component, a reverse shock SSC
component or a cross IC component(forward/reverse shock synchrotron
photons scattered by electrons from reverse-shocked/forward-shocked
region), a high
latitude (curvature) component of the prompt high energy emission.

\subsection{Reverse Shock Synchrotron Emission}

We consider the thin shell case as is assumed in \S 3, the flux
peaks at the crossing time of the reverse shock $t_{\times}=t_{\rm
dec}>T$ with Lorentz factor $\Gamma_{\times}=\Gamma_{\rm d}$, and the
ratio of the comoving number densities of the forward-shocked to
that of the reverse-shocked regions is given by ${n_{\rm r}}/{n_{\rm
f}}\sim\Gamma_{\rm d}\sim 2.3\times 10^3
E_{53.5}^{1/8}n_{-4}^{-1/8}t_{\rm dec}^{-3/8}$
\citep{Kobayashi2007}. The internal energy densities $e$ and the
bulk Lorentz factors $\Gamma$ of the two regions are equal with each other \citep{Zhang2003}.
Consequently, we have that
\begin{equation}
\frac{\nu_{\rm m}^{\rm r}}{\nu_{\rm m}^{\rm f}}=\left(\frac{n_{\rm r}}{n_{\rm f}}\right)^{-2}\Re_B\Re_e^2, ~~
\frac{\nu_{\rm c}^{\rm r}}{\nu_{\rm c}^{\rm f}}=\Re_B^{-3}\left[1+Y(\gamma_{c,\times}^{\rm r})\right]^{-2}, 
~~\frac{F_{\rm max}^{\rm r}}{F_{\rm max}^{\rm f}}=\frac{n_{\rm r}}{n_{\rm f}}\Re_B,
\end{equation}
characterized by the ratios $\Re_B\equiv\left(\frac{\epsilon_{B\rm r}}{\epsilon_{B\rm f}}\right)^{\frac{1}{2}}$, 
$\Re_e\equiv\frac{\epsilon_{e\rm r}}{\epsilon_{e\rm f}}$.

The minimum and cooling frequencies of the reverse shock synchrotron emission at the crossing time under the 
low energy constraints are
\begin{equation}
h\nu_{\rm m,\times}^r=1.1\epsilon_{e\rm f,-0.4}E_{53.5}^{-1/4}t_{\rm dec}^{-3/4}\Re_{B,1}\Re_{e,0}^{2}
g_pg_{e2}^{4.0}{\rm eV},
\end{equation}
\begin{equation}
h\nu_{\rm c,\times}^r=3.9\times 10^7\epsilon_{e\rm f,-0.4}^{2}E_{53.5}^{2}t_{\rm dec}^{-1/2}\Re_{B,1}^{-3}
g_p^{2}g_{e2}^{-10}\left[1+Y(\gamma_{c,\times}^{\rm r})\right]^{-2} {\rm eV}.
\end{equation}

According to \S 2.2, we have 
$Y(\gamma_*^{\rm r})<Y(\gamma_c^{\rm r})\ll1$ around the deceleration time, thus the 
synchrotron spectrum does not change and the ratio of the reverse shock synchrotron 
flux to forward shock synchrotron flux can be calculated as
\begin{equation}\label{fratio}
\frac{F_{1 \rm GeV,\times}^{\rm r}}{F_{1\rm GeV,\times}^{\rm f}}=\frac{F_{\rm max,\times}^{\rm r}}{F_{\rm 
max,\times}^{\rm f}}\left(\frac{\nu_{\rm m,\times}^{\rm r}}{\nu_{\rm m,\times}^{\rm f}}\right)^{\frac{p-1}
{2}}\left(\frac{h\nu_{\rm c,\times}^{\rm r}}{1 \rm GeV}\right)^{\frac{1}{2}}=\left(\frac{n_{\rm r}}{n_{\rm f}}
\right)^{-p+2}\Re_{e}^{p-1}\Re_{B}^{(p+1)/2}\left(\frac{h\nu_{c,\times}^{\rm r}}{1\rm GeV}\right)^{1/2},
\end{equation}
Inserting equations (\ref{eBcon}) and (\ref{ncon}) into equation (\ref{fratio}),
the ratio turns to be
\begin{equation}
\frac{F_{1 \rm GeV,\times}^{\rm r}}{F_{1\rm GeV,\times}^{\rm f}}=0.23\epsilon_{e\rm f,-0.4}^{p/2}E_{53.5}^{-
p/4+3/2}t_{\rm dec}^{3p/8-1}\Re_{B,1}^{p/2-1}\Re_{e,0}^{p-1}g_{e1}^{-6.7}g_p^{p/2}g_{e2}^{-p-3.0}.
\end{equation}
The ratio could be as high as $0.59$ since the total kinetic energy can be as high as 
$E\sim9.0\times10^{53}{\rm erg}$ by considering a reasonable radiation efficiency $\gtrsim10\%$.
Considering equation (\ref{F_f}), the flux density from reverse shock synchrotron emission cannot be as high as 
$\sim\rm\mu Jy$ to explain the observations at early times.
Thus, the reverse shock synchrotron emission is unlikely to contribute to the 
early-time LAT observation.

\subsection{The SSC and EIC Emission}

Besides the reverse synchrotron emission, we consider the other four possible IC processes, including the 
synchrotron self-Compton (SSC) processes in forward and reverse shocks, and two combined-IC processes (i.e. 
scattering of reverse-shock synchrotron photons on electrons accelerated in forward shocks and froward-shock 
synchrotron  photons on electrons accelerated in reverse shocks).

The optical depth of inverse Compton scattering in the reverse shock in the Thomson regime is
$\tau^{\rm r}=\frac{1}{3}\sigma_{\rm T} n_{\rm r} R_{\rm d}/\Gamma_d=7.7\times 10^{-8}
\epsilon_{e\rm f,-0.4}^{5/2}E_{53.5}^{-1/4}t_{\rm dec}^{-1/8}g_p^{5/2}g_{e2}^{-5.0}$\citep{Wang2001}.
The optical depth to inverse Compton scattering in the forward shock in the Thomson regime is
$\tau^{\rm f}=\frac{1}{3}\sigma_{\rm T} n R_{\rm d}=3.8\times 10^{-12}
\epsilon_{e\rm f,-0.4}^{3}E_{53.5}^{-1/2}t_{\rm dec}^{1/4}g_p^{3}g_{e2}^{-6.0}$.
Taking the peak flux of synchrotron emission at the crossing time of the
reverse shock and forward shock, which are $F_{\rm max}^{\rm r}=0.30\epsilon_{e\rm f,-0.4}^{-1/2}E_{53.5}^
{1/4}
t_{\rm dec}^{-3/8}\Re_{B,1}g_p^{-1/2}g_{e2}^{3.0}{\rm Jy}$ and
$F_{\rm max}^{\rm f}=12 g_{e2}^{2.0} {\rm \mu Jy}$, respectively, we can get
the peak flux density of the four Inverse-Compton components as follows \citep{Wang2001},
\begin{equation}
F_{\rm max}^{\rm IC,rr}=\tau^{\rm r}F_{\rm max}^{\rm r}=2.2\times 10^{-8}{\rm Jy}\epsilon_{e\rm f,-0.4}^{2}t_
{\rm dec}^{-1/2}\Re_{B,1}g_{p}^2g_{e2}^{-2.0},
\end{equation}
\begin{equation}
F_{\rm max}^{\rm IC,rf}=\tau^{\rm r}F_{\rm max}^{\rm f}=8.8\times 10^{-13}{\rm Jy}\epsilon_{e\rm f,-0.4}^
{5/2}E_{53.5}^{-1/4}t_{\rm dec}^{-1/8}g_p^{5/2}g_{e2}^{-3.0},
\end{equation}
\begin{equation}
F_{\rm max}^{\rm IC,fr}=\tau^{\rm f}F_{\rm max}^{\rm r}=1.1\times 10^{-12}{\rm Jy}\epsilon_{e\rm f,-0.4}^
{5/2}E_{53.5}^{-1/4}t_{\rm dec}^{-1/8}\Re_{B,1}g_p^{5/2}g_{e2}^{-3.0}
\end{equation}
and
\begin{equation}
F_{\rm max}^{\rm IC,ff}=\tau^{\rm f}F_{\rm max}^{\rm f}=4.5\times 10^{-17}{\rm Jy}\epsilon_{e\rm f,-0.4}^{3}
E_{53.5}^{-1/2}t_{\rm dec}^{1/4}g_p^{3}g_{e2}^{-4.0},
\end{equation}
respectively, where the superscripts `rr' and `ff' mean the SSC
emission in reverse shock and forward shock respectively, `fr' and
`rf' mean the scattering of reverse shock photons on the electrons
in forward shocks and forward shock photons on the electrons in
reverse shocks. Therefore, the IC contributions (except for the SSC emission 
in the reverse shock) can be excluded, since even their peak fluxes are much smaller
 than the early-time LAT observations. Although the peak flux of the SSC emission 
in the reverse shock is close to the observed flux,  the reverse shock SSC emission 
can also be excluded because its flux peaks at a low energy of about $20 {\rm keV}$, 
indicating that the flux in the LAT band is about $6$ orders lower than the peak flux. 
A low flux for these IC processes is mainly due to a very low circum-burst density 
in GRB 090510 inferred from the low-energy observations.

\subsection{The High Latitude Prompt Emission}

The spectrum of GRB 090510 shows a power-law component in the LAT band
($>100{\rm MeV}$), whose physical origin at early times is unclear.
One might consider the early part of the extended emission (shown by the gray dotted line with {a decay slope
$\alpha_{\rm Lat}(0.37-3{\rm s})=2.0$} in Figures 1 and 2) is due to the high-latitude emission of the prompt 
emission \citep{Kumar2000}.
Because photons from high latitude regions with respect to the line of sight will
arrive later than that from low latitude region due to the curved front surface of the jet,
one observes a fast decreasing emission rather than an abrupt stop of the emission,
the so-called  ``curvature effect".
Then the high latitude emission flux of the prompt emission evolves as
\begin{equation}\label{cur}
F_{\nu}(t) \propto\left[\frac{t-(t_0-\Delta t)}{\Delta t}\right]^{-2-\beta}
\end{equation}
according to \citet{Toma2009cocoon}(The details are shown in Appendix B), where
$t_0$ is the pulse peak time. 
According to \citet{Ackermann090510} and \citet{Abdo090510}, the variability timescale 
derived from the BGO emission ($100{\rm keV}-$few MeV) is about $14\pm2{\rm ms}$ 
in the time interval $0.17-0.37{\rm s}$ 
since the BAT trigger, but the variability timescale for the LAT emission is not 
determined.  If we assume that the variability timescale for the high energy 
( larger than $100{\rm MeV}$) emission, which is dominated by the power law component, 
is the same as that of BGO observation, then we adopt $\Delta t\sim0.01{\rm s}$ 
in equation (\ref{cur}).
Thus, the high latitude flux of the prompt emission at $t_0+0.1 {\rm s}$ is 
$F(t_0+0.1 {\rm s})=11^{-2-\beta}F(t_0)$. 
Since the index of the power law component in the time interval $0.17-0.37{\rm s}$ 
is $1.66\pm0.04$ \citep{Ackermann090510, Abdo090510}, we take the rough value 
$\beta\sim0.66$. 
Then the flux of the high-latitude emission at $t_0+0.1{\rm s}$ will be reduced by 
a factor of $\sim600$ relative to that of the line-of-sight emission at $t_0\sim 0.37$, 
which is too steep to explain the early high energy emission at $t=0.37-3{\rm s}$.

Since the origin of the LAT emission in the prompt phase, 
dominated by the extra power-law component, is unclear 
and can be different from that of the Band component in 
the GBM energy range, the variability timescale in the 
LAT range can generally be different from $\Delta t \sim 0.01{\rm s}$. 
In particular, the emission radius of the extra high energy 
component may be much larger than that of the Band component, 
which may lead to a much larger pulse width in the LAT energy 
range than in the GBM energy range (see e.g., \citet{Toma2010}).
If we take $\Delta t\sim 0.5{\rm s}$, then the high latitude 
flux of the prompt emission at $t_0\sim0.37{\rm s}$ evolves as 
$F(t)\sim(2t+0.26{\rm s})^{-2.66}F(0.37{\rm s})$, which together with the synchrotron 
forward shock emission with a deceleration time $t_{\rm dec}\leq(2-3){\rm s}$ can 
explain the early high energy emission at $t\sim0.37-3{\rm s}$, shown in Fig 1.

\section{The Line-of-Sight Prompt Emission as the Origin of the
Early-Time Afterglow High-Energy Emission}

As a final possibility, we consider the possible influence of the tail-end 
of the prompt emission in the description of the early afterglow.
In \S 3 and \S 4 we assumed a small GRB duration $T_{90}=0.30\pm0.07{\rm s}$,
following \citet{DePasquale2010} based on the observations in the  GBM and
BAT bands. Under this assumption, we discussed the forward shock synchrotron emission,
the forward reverse shock synchrotron emission, the four possible crossed inverse
Compton processes, and the high latitude emission of the prompt emission, and we concluded
that none of them can explain the early decaying part of the  high energy emission at
$0.3\lesssim t \lesssim 3{\rm s}$ except for the high latitude portion of 
the prompt emission with a larger variability time scale plus the synchrotron forward 
shock emission with a deceleration time $t_{\rm dec}\leq(2-3){\rm s}$.

We focus on the BAT detections around $T_0 + (1-3)$ s, 
shown in \citet{DePasquale2010}, which implies that 
the prompt emission could last until such times. We 
calculate the flux of the synchrotron forward shock emission
at $t=3$s for $p=2.5$ in the BAT energy range, and obtain 
$\sim 52 {\rm \mu Jy}$ (the flux is similar for $p=2.2$). This is about
one order of magnitude smaller than the observed flux.
Thus these BAT detections are not from the external shock,
but may be attributed to the prompt emission. 

The LAT prompt emission may also last until such times. 
\citet{Ackermann090510} obtain no significant temporal 
correlation between NaI ($8{\rm keV}-260{\rm keV}$) data and the LAT data 
$>100{\rm MeV}$ between $t=0.17$ s and $t=0.47$ s. This does not 
necessarily indicate that the LAT prompt emission
(the extra component) and the GBM prompt emission (Band
component) have fully unrelated origins. Some theoretical
models suggest that the two components may arise at 
different radii (e.g., \citet{Toma2010}, \citet{Wang2006}) 
in the same shells, or the two components may arise by the 
leptonic and hadronic processes in the same internal shock 
(e.g., \citet{Asano2009}; \citet{Razzaque2010}). Therefore
it appears reasonable to consider that GRB 090510 has 
a duration as long as $2-3$ s and the high energy emission 
before $t\sim 2-3$ s is a part of the prompt emission.

Under this new assumption,  with a longer GRB duration $T\sim2.0{\rm s}$, one infers a
critical Lorentz factor $\Gamma_c\sim 2.0\times 10^3 E_{53.5}^{1/8}n_{-3}^{-1/8}
T_{0.3}^{-3/8}$. For the thin shell case with an initial Lorentz factor $\Gamma_0$
smaller than $\Gamma_c$, the shell will be decelerated  efficiently by the reverse shock
after the GRB main duration. The high energy light curve with the initial Lorentz factor
$\Gamma_0=1900$ and the number density $n=10^{-3} {\rm cm^{-3}}$ is shown by the thin
solid line in Figure 2, which can explain the late-time high energy emission well.
If we assume that the LAT emission around $t=2{\rm s}$ is a flare of duration
$\Delta t=1{\rm s}$, which peaks at $t_0=2{\rm s}$ with a peak flux $F(t_0)$, and 
having \citep{DePasquale2010} a spectral index $\beta=0.41^{+0.28}_{-0.31}$ at 
$t=1.5-2.5{\rm s}$, then the flux due to curvature effect at $t=3.5{\rm s}$ is 
$F(t)=0.11F(t_0)$, which is compatible with the seventh data point.
In this case, the deceleration time could in principle be even larger, 
implying a smaller Lorentz factor, but due to the lack of precise information 
about the last flare, e.g., its duration and the flux peak time, we cannot use this 
to derive a firmer deceleration time.

For a larger initial Lorentz factor $\Gamma_0>\Gamma_{c}$, the reverse shock will
transition from the Newtonian phase to the relativistic phase at the time
$t_{\rm N}=\left({E}/{16\pi \Delta n m_p c^4\Gamma_0^8}\right)^{1/2}=0.081E_{53.5}^{1/2}
\Gamma_{0,3.7}^{-4}n_{-4}^{-1/2}{\rm s}$ \citep{Sari1997dyn}, 
with the width of the shell in the comoving frame
$\Delta=cT/(1+z)=3.1\times 10^{10} T_{0.3} {\rm cm}$. The deceleration time when the
dissipated energy is comparable to the total kinetic energy of the shell is about
$t_{\rm dec}=2T$. At the time $t_{\rm N}<t<t_{\rm dec}$,  the Lorentz factor of the forward
shock evolves as $\Gamma\propto t^{-1/4}$, and the radius of the shell evolves as
$R\propto t^{1/2}$. Thus, the characteristic frequencies and the peak flux density evolve as
$\nu_{\rm min}^{\rm r}\propto t^{-1}$, $\nu_{\rm c}^{\rm r}\propto t^{-1}$ and
$F_{\rm max}\propto t$. As a result, the flux of high energy emission in the region
$\nu_{\rm m}^{\rm f}<\nu_{\rm LAT}<\nu_{\rm c}^{\rm f}$ evolves as
$F_{\rm LAT}\propto t^{(-p+3)/2}$; and the flux density of high energy emission in the region
$\nu_{\rm LAT}>\nu_{\rm c}^{\rm f}$ evolves as $F_{\rm LAT}\propto t^{(-p+2)/2}$.
This shows  that the forward shock synchrotron emission in the thick shell case cannot
explain the early-time high energy emission due to its flatter light curve.
And the reverse shock emission in the thick shell case can also be excluded, using
similar calculations as in \S 4.

Based on the above analysis, while the forward shock synchrotron afterglow dominates the LAT
emission at late times, the early times LAT emission must be attributed to a prompt
emission, which can be limited from above by a high-latitude curvature emission envelope.

\section{Discussion and Conclusions}
\label{sec:disc}

We have investigated whether the photons received by the LAT can be explained solely by
the forward shock afterglow emission or whether the prompt emission necessarily contributes
to the early part of this emission. We have addressed this question with the help of the
broad-band observations from the XRT and UVOT instruments onboard of Swift and the LAT  
onboard of Fermi for the case of the short GRB 090510, obtaining constraints on the 
extended high-energy emission based on the XRT and UVOT observations.

We have obtained a good fit to the XRT and UVOT observations 
{(except for the late-time optical data)} 
in terms of synchrotron 
emission from an adiabatic forward shock in spreading jet model, which constrains the 
environment of this short burst, implying a low number density $n \sim 10^{-3}- 10^{-6}
{\rm cm^{-3}}$, consistent with a binary progenitor scenario 
\citep[e.g.,][]{Belczynski2006}, corresponding to an electron energy fraction 
$\epsilon_{e\rm f} \sim 0.6 - 0.2$ and a weak magnetic field with an energy fraction
$\epsilon_{B\rm f}\sim 10^{-5}- 10^{-2}$. The latter could in principle be due to an 
upstream magnetic field of $\sim 7\mu$G just shock-compressed, without need for magnetic 
field amplification \citep{Kumar20093GRBs}. However, the magnetic field values of 
$\sim \mu$G derived in our galaxy correspond to an average ISM of density $n\sim 1$ 
cm$^{-3}$, whereas the fits for GRB 090510 indicate external densities comparable to 
a halo or intergalactic medium with $n \sim 10^{-3} - 10^{-6} {\rm cm^{-3}}$, where 
based on flux-freezing the field would be expected to be much less than $\mu$G. 
Thus, the magnetic field may still need to be amplified in the shock.  We note also that 
while the electron energy fraction can be as high as $0.6$, the adiabatic condition is 
still satisfied, since the low energy emission constrains the radiation to be in the 
slow cooling case, indicating a radiation efficiency much smaller than unity,
in contrast to the radiative regime assumption of \citet{Ghisellini2010}, where they did not
take into account low energy constraints.

In \citet{DePasquale2010}, it is assumed 
that the LAT data at $t=1-200s$ with slope $-1.38$ arises from synchrotron 
emission in the regime $\nu_{\rm LAT}>\nu_{\rm c}^{\rm f}$, i.e., $\nu_c<10^8{\rm eV}$,
in contrast to our conclusion that the synchrotron emission in the regime
$\nu_{\rm c}^{\rm f}>1{\rm GeV}$ evolving with slope $\sim-1.1$. 
Part of the difference may be due to \citet{DePasquale2010} using the formulae 
of \citet{Granot2002}, while we used the formulae of 
\citet{Sari1998afterglow}.
However, applying the constraints on $\epsilon_B$,$n$, $\epsilon_{e\rm f}E$ in 
\citet{DePasquale2010} and taking $p=2.5$ in the expression for 
$\nu_{\rm c}^{\rm f}$ from \citet{Granot2002}, one obtains
$\nu_{\rm c}^{\rm f}\simeq 6.3\times10^7{\rm eV}$ at $100$ s.  Therefore, 
$\nu_{\rm c}^{\rm f}=6.3\times10^8{\rm eV}$ at $1{\rm s}$ since $\nu_c\propto t^{-1/2}$,
falling in the LAT band,  inconsistent with the \citet{DePasquale2010} assumption. 
The simple power-law decay shown in their Figure 1 is not applicable
due to the $\nu_c$ crossing through the LAT band, and it overestimates the LAT 
photon flux by a factor of $1.7$ at $t=1 {\rm s}$ and $2.1$ at $t\sim 0.4{\rm s}$ . 
We note also that the formulae in \citet{Granot2002} 
is applicable only in the self-similar phase, which takes a few e-folding
times after the deceleration time to fully develop.
Our numerical calculation shows that around the deceleration time, the flux should
be reduced by a factor of $\sim2-3$.
Furthermore, if the deceleration time is not larger than the duration $T$, 
the LAT light curve will be flatter at $t<T$, which cannot explain the 
early-time LAT emission, as discussed
in \S 5 and also strongly argued by \citet{Maxham2011}.

The initial jet angle is found to be constrained to be $\theta_{\rm j} \lesssim 1^{\circ}$, which
implies a collimation-corrected kinetic energy in the range of short GRBs.  We have used for
our numerical calculations the semi-analytical jet spreading model of \citet{Huang2000apj}
(detailed in Appendix A), assuming that the observer line of sight is approximately
along the jet axis.   This leads us to identify a feature in the late XRT and UVOT light
curves as symptomatic of a jet break. 
We note however that \citet{VanEerten2010} argues that numerically the light curve
break related to a jet edge as seen by an off-axis observer may appear hidden for a 
longer time (weeks) than it would be for an on-axis observer, and the alternative 
possibility cannot be ruled out, although this being one of the brightest bursts 
reduces the probability of its being observed from an off-axis direction.

We have considered the origin of the high energy emission under the constraints provided by
the low energy observations. Calculation of the Klein-Nishina effect on the IC scattering of
synchrotron photons by electrons accelerated in the forward shock indicate that the 
Klein-Nishina
effect suppresses the high energy IC emission significantly, which  results in a Y-parameter
much smaller than unity. The distribution of the high energy electrons is not affected by
the Klein-Nishina effect, so that the synchrotron high energy emission cannot steepen due to
a suppression of the IC scattering through Klein-Nishina effects as argued in \citet{Wang2010}.

The duration of the prompt emission of this GRB is a crucial parameter for establishing the
physics of the LAT emission. Based on the phenomenon that the Fermi GBM emission
turns over sharply at $t\sim 0.30\pm 0.07{\rm s}$ \citep{DePasquale2010},  we
adopted a duration of $T=0.30\pm 0.07{\rm s}$, compatible with a thin shell case where the
deceleration time is larger than the GRB duration. Here, however, we find that a forward 
shock synchrotron emission model agrees well only with the late-time ($t>3{\rm s}$) 
high-energy observations of the LAT, but this model falls well below the early-time 
high-energy emission (the first six LAT data points). It cannot contribute to the 
power-law high energy component in the prompt spectrum, which suggests that the 
early-time high-energy emission may have a different origin from the late-time 
high-energy emission. We excluded various other possibilities for the early-time 
high-energy emission, such as synchrotron emission from the reverse external shock, 
SSC emission from reverse/forward external shock, and IC scattering of forward/reverse 
synchrotron photons by electrons accelerated in reverse/forward shocks, as well as the 
high-latitude prompt emission with a small variability timescale. 
The latter is too steep to contribute to the early-time high-energy emission, due 
to the assumed high temporal variability of the prompt emission,
unless  the early-time high-energy emission variability timescale is as large as $0.5$ s.

We are led to the conclusion that the early-time high-energy emission is likely
due to the final portions of the prompt component, or else to the high latitude 
component of the prompt emission with a variability time scale as large as $0.5$ s 
plus the synchrotron emission with a deceleration time $t_{\rm dec}\leq(2-3){\rm s}$.  
This is supported by
the fact that we cannot explain the first six LAT data points by synchrotron 
forward shock emission if we consider the combined data in the three bands 
(XRT, UVOT and LAT).
In this case, the long-lived high-energy emission can be naturally explained 
as the result of the prompt emission at early times ($t<(2-3) {\rm s}$), and the 
afterglow emission at late times ($t>(2-3) {\rm s}$) from the adiabatic forward 
shock synchrotron radiation, with reasonable parameters. In this two-component 
model of the long-lived high energy emission, the shock deceleration time 
of $t_{\rm dec}=2{\rm s}$ results in a fit with an initial Lorentz 
factor around $\Gamma\simeq 2000$,
not much larger than the lower limits $\Gamma_{\rm min}
=950\pm40$ and $1220\pm60$ \citep{Abdo090510} obtained from the pair-production limits 
implied  by the presence of $3.4 \rm GeV$ and $31 \rm GeV$ photons.
With smaller values of $t_{\rm dec}$, very large initial Lorentz factors are
needed (but see \citet{Ioka2010}). 
Although there is so far no statistically significant evidence for an early 
steeper decay, due to the sparse nature of the LAT data, it is striking that by eye
there are 4 out 10 GRB long-lived LAT light curves which have relatively steeper slopes
than the slopes, $\sim -1$, of typical late-time X-ray and optical afterglows, 
according to Table 3 and the light curve fits in the figures of \citet{Zhang2010catalog}.
The steep decay slope of the LAT light curve is difficult to explain with a normal 
external shock model. A similar conclusion was subsequently suggested by \citet{Liu2010},
and also by \citet{Maxham2011}.
Our two-component model also explains the steep high-energy temporal decay index 
values $\alpha_{\rm LAT} \sim 1.5$ as being the natural result of the superposition
of the tail-end of the prompt regime and the start of the afterglow. 
While our model for the high energy afterglow is in principle applicable to other LAT 
bursts as well, a homogeneous sample of similarly detailed data including low energy 
constraints will be needed, on a large number of objects, before definite conclusions 
can be reached.

\acknowledgements{We thank D. Burrows, D. Fox, M. De Pasquale, R. Barniol Duran, 
S. Razzaque, B. Zhang, B.B. Zhang and Y. Z. Fan for useful discussions; Dr. De Pasquale for 
supplying optical data to us; and the referee for useful comments and suggestions.
Partial support was provided by NASA NNX08AL40G, NNX09AT72G, NSF PHY0757155, the NSFC 
under grants 10973008, the 973 program under grants 2009CB824800, National Natural 
Science Foundation of China grant 11033002, the Foundation for the Authors of National 
Excellent Doctoral Dissertations of China, the Program for New Century Excellent
Talents in University,  the Qing Lan Project and the Fok Ying Tung Education 
Foundation and China Scholarship Council Postgraduate Scholarship Program.}

\appendix
\section{Jet Deceleration and Spreading Dynamical Model }

We solve the following equations for the evolution of the shock radius $R$, the mass $m$ swept
up by the shock, the opening half-angle of jet $\theta$ and the Lorentz factor of
shock $\Gamma$ \citep[e.g.,][]{Huang2000apj}.

The evolution of the shock radius  is described by
\begin{equation}
\frac{dR}{dt}=\beta c\Gamma(\Gamma+\sqrt{\Gamma^2-1}),
\end{equation}
with $t$ as the observer's time and $\beta=\sqrt{\Gamma^2-1}/\Gamma$.
The swept mass evolves as 
\begin{equation}
\frac{dm}{dR}=2\pi R^2(1-\cos\theta)nm_p,
\end{equation}
with $m_p$ as the proton mass and $n$ as the number density of surrounding medium.
The evolution of the opening angle considering the jet spreading is described as
\begin{equation}
\frac{d\theta}{dt}\equiv\frac{1}{R}\frac{da}{dt}=\frac{c_{\rm s}(\Gamma+\sqrt{\Gamma^2-1})}{R},
\end{equation}
where $a$ is the comoving lateral radius of the jet \citep{Rhoads1999, Moderski2000}
and  the comoving sound speed
$c_{\rm s}=\hat{\gamma}(\hat{\gamma}-1)(\Gamma-1)\frac{1}{1+\hat{\gamma}(\Gamma-1)}c^2$
with the adiabatic index $\hat{\gamma}$ as $\hat{\gamma}\approx(4\Gamma+1)/(3\Gamma)$.
The conservation of the total energy can be expressed as
\begin{equation}
\frac{d\Gamma}{dm}=-\frac{\Gamma^2-1}{M_{\rm ej}+\eta_{\rm r} m+2(1-\epsilon)\Gamma m},
\end{equation}
where the radiative efficiency is $\eta_{\rm r} =0$ for the adiabatic case,  and
$M_{\rm ej}$ is the ejecta mass.

\section{The High-latitude emission}

Because photons from high latitude region with respect to the line of sight will arrive later than that from 
low latitude region due to the curved front surface of the jet, 
we will observe a fast decreasing emission instead of an abrupt cutoff of the emission, 
which is so called ``curvature effect".

We neglect the radial structure of the emitting shell for simplicity. 
Under this assumption we calculate the flux from the shell at radius $r_i$ which expands toward us 
with Lorentz factor $\Gamma_i$ by
 \citep{Granot1999, Woods1999, Ioka2001, Yamazaki2003, Dermer2004, Toma2009cocoon}
\begin{equation}
F_{\nu}(t) = \frac{1+z}{d_{\rm L}^2} 8\pi r_i^3 j'_{\nu'} \frac{1}{[1+\Gamma_j^2 \theta^2(t)]^2},
\end{equation}
where the photon frequency in comoving frame is
\begin{equation}
\nu' = (1+z)\nu \frac{1+\Gamma_j^2 \theta^2(t)}{2\Gamma_j}
\end{equation}
with $\nu$ as the photon frequency in observing frame, and $\theta(t)$ describes the emitting point of the 
high-latitude emission.
We have the comoving emissivity
 \begin{equation}
j'_{\nu'}\propto \nu'^{-\beta} \propto (1+\Gamma_j^2 \theta^2(t))^{-\beta} ,
\end{equation}
where $\beta$ is the spectral index of the observed emission, which is consistent with the power-law spectra fit 
of \citep{Abdo090510}.
By using the relationship
$\theta(t) = \sqrt{2} \left[1- \frac{c}{r_i}
\left(\bar{t}_i - \frac{t}{1+z}\right)\right]^{1/2}$
with $\bar{t}_i$ as the emission time in the central engine frame,
we get
\begin{equation}
1+\Gamma_j^2 \theta^2(t)=\frac{t-(t_0-\Delta t)}{\Delta t}
\end{equation}
where $t_0=(1+z)(\bar{t}_i-\frac{r_i}{c})$ is the observed peak time of the high energy pulse
 and $\Delta t=\frac{r_i(1+z)}{2c\Gamma_i^2}$ is the observed dynamical time scale.
Therefore, the observed flux evolves as time as following
\begin{equation}
F_{\nu}(t) \propto\left[\frac{t-(t_0-\Delta t)}{\Delta t}\right]^{-2-\beta}.
\end{equation}

\clearpage

\begin{figure}
\plotone{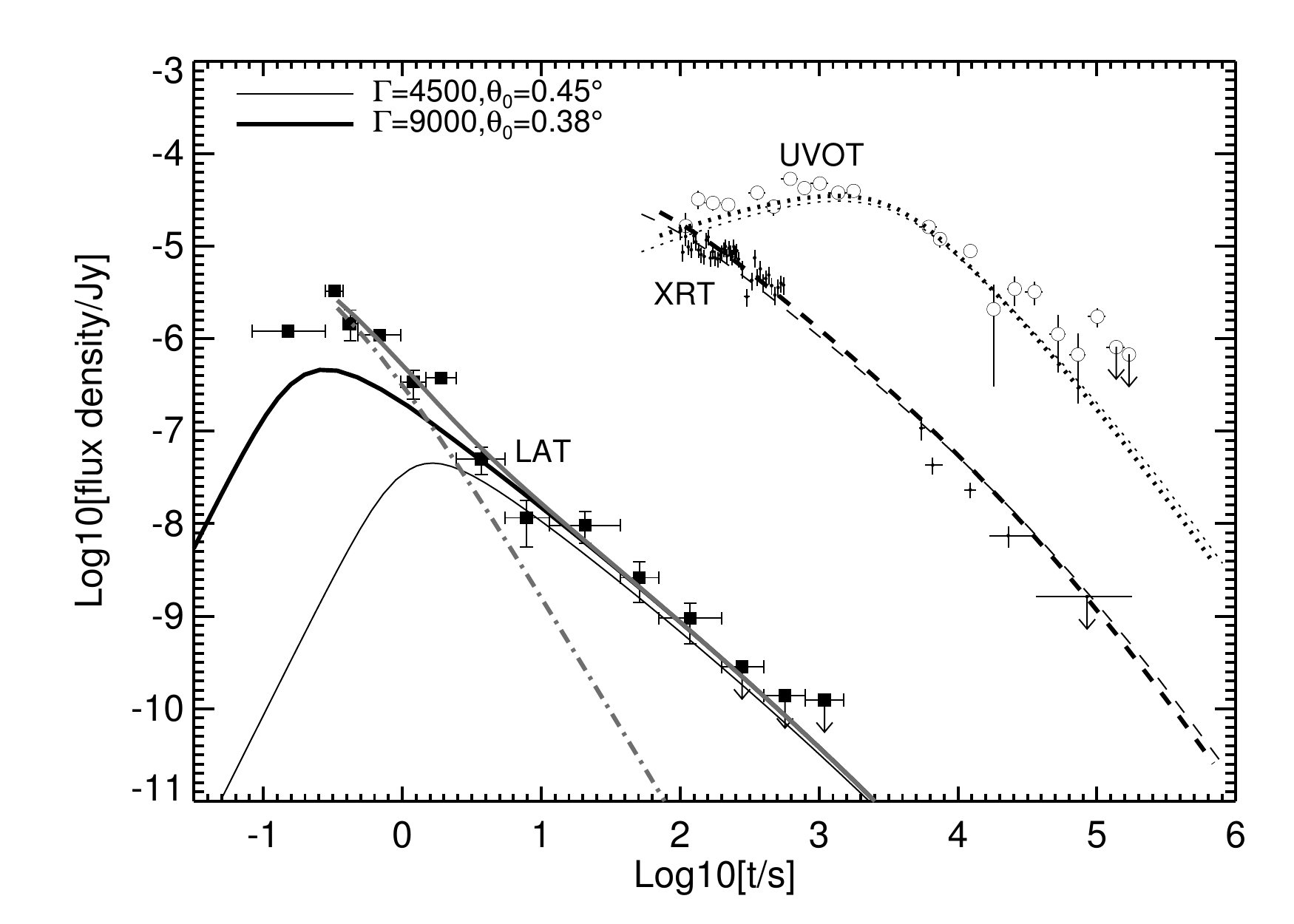}\caption{GRB 090510 broad-band fitting.  
The data are the energy flux densities averaged in the observed energy bands 
\citep{DePasquale2010}: LAT(100MeV-4GeV, filled squares); XRT (0.2 keV - 10 keV,  
crosses); UVOT (white band, open circles). The black dotted, dashed and solid lines 
are the light curves of the optical, X-ray and high energy emission, respectively, 
from the adiabatic forward shock synchrotron emission in the jet spreading 
model with parameters $E=3\times10^{53} \rm{ergs}$, $\epsilon_e=0.17$, 
$\epsilon_B=7\times10^{-3}$, $n=1\times10^{-6}{\rm cm^{-3}}$ and electron 
index $p=2.5$. The thick lines are for an initial Lorentz factor $\Gamma_0=9000$ and
an initial jet half-angle $\theta_0=0.38^{\circ}$;  the thin lines are for 
$\Gamma_0=4500$ and $\theta_0=0.45^{\circ}$.  The thick gray dotted line shows 
a hypothetical component, which can be explained by the high latitude emission of 
the prompt emission with a variability timescale $\Delta t=0.5{\rm s}$, which
makes up for the difference between the synchrotron emission with $\Gamma_0=9000$ 
and the observations. The thick gray solid line is the sum of the thick gray 
dotted line and the thick black solid line.
}
\end{figure}

\begin{figure}
\plotone{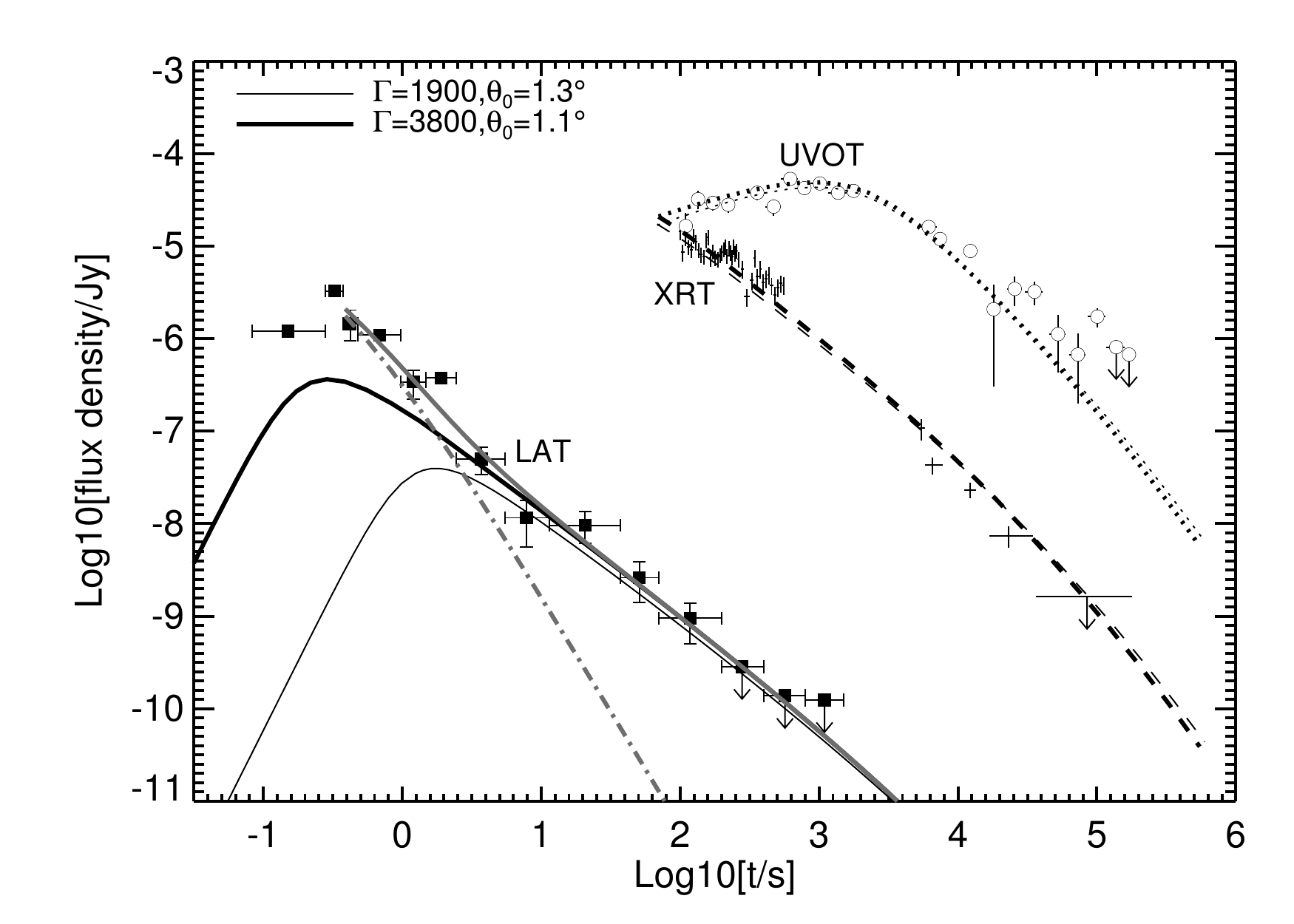}\caption{
GRB 090510 broad-band fitting via the adiabatic forward shock synchrotron emission 
in the jet spreading model, with parameters $E=3\times10^{53} \rm{ergs}$, 
$\epsilon_e=0.6$, $\epsilon_B=10^{-5}$, $n=10^ {-3}{\rm cm^{-3}}$ and $p=2.5$.  
The thick lines are for an initial Lorentz factor $\Gamma_0=3800$ and an
initial jet half-angle $\theta_0=1.3^{\circ}$, while the thin lines are for 
$\Gamma_0=1900$ and $\theta_0=1.1^{\circ}$.  The thick gray dotted line shows 
a hypothetical component, which can be explained by the high latitude emission 
of the prompt emission with a variability timescale $\Delta t=0.5{\rm s}$, which
makes up for the difference between the synchrotron emission with $\Gamma_0=3800$ 
and the observations.  The thick gray solid line is the sum of the thick gray 
dotted line and the thick black solid line. 
}
\end{figure}


\clearpage


 \begin{thebibliography}{59}
\expandafter\ifx\csname natexlab\endcsname\relax\def\natexlab#1{#1}\fi

\bibitem[{{Abdo} {et~al.}(2009{\natexlab{a}}){Abdo}, {Ackermann}, {Ajello},
  {Asano}, {Atwood}, {Axelsson}, {Baldini}, {Ballet}, {Barbiellini}, {Baring},
  {Bastieri}, {Bechtol}, {Bellazzini}, {Berenji}, {Bhat}, {Bissaldi},
  {Blandford}, {Bloom}, {Bonamente}, {Borgland}, {Bouvier}, {Bregeon}, {Brez},
  {Briggs}, {Brigida}, {Bruel}, {Burgess}, {Burrows}, {Buson}, {Caliandro},
  {Cameron}, {Caraveo}, {Casandjian}, {Cecchi}, {{\c C}elik}, {Chekhtman},
  {Cheung}, {Chiang}, {Ciprini}, {Claus}, {Cohen-Tanugi}, {Cominsky},
  {Connaughton}, {Conrad}, {Cutini}, {d'Elia}, {Dermer}, {de Angelis}, {de
  Palma}, {Digel}, {Dingus}, {Silva}, {Drell}, {Dubois}, {Dumora}, {Farnier},
  {Favuzzi}, {Fegan}, {Finke}, {Fishman}, {Focke}, {Fortin}, {Frailis},
  {Fukazawa}, {Funk}, {Fusco}, {Gargano}, {Gehrels}, {Germani}, {Giavitto},
  {Giebels}, {Giglietto}, {Giordano}, {Glanzman}, {Godfrey}, {Goldstein},
  {Granot}, {Greiner}, {Grenier}, {Grove}, {Guillemot}, {Guiriec}, {Hanabata},
  {Harding}, {Hayashida}, {Hays}, {Horan}, {Hughes}, {Jackson},
  {J{\'o}hannesson}, {Johnson}, {Johnson}, {Johnson}, {Kamae}, {Katagiri},
  {Kataoka}, {Kawai}, {Kerr}, {Kippen}, {Kn{\"o}dlseder}, {Kocevski}, {Komin},
  {Kouveliotou}, {Kuss}, {Lande}, {Latronico}, {Lemoine-Goumard}, {Longo},
  {Loparco}, {Lott}, {Lovellette}, {Lubrano}, {Madejski}, {Makeev},
  {Mazziotta}, {McBreen}, {McEnery}, {McGlynn}, {Meegan}, {M{\'e}sz{\'a}ros},
  {Meurer}, {Michelson}, {Mitthumsiri}, {Mizuno}, {Moiseev}, {Monte},
  {Monzani}, {Moretti}, {Morselli}, {Moskalenko}, {Murgia}, {Nakamori},
  {Nolan}, {Norris}, {Nuss}, {Ohno}, {Ohsugi}, {Omodei}, {Orlando}, {Ormes},
  {Paciesas}, {Paneque}, {Panetta}, {Pelassa}, {Pepe}, {Pesce-Rollins},
  {Petrosian}, {Piron}, {Porter}, {Preece}, {Rain{\`o}}, {Rando}, {Rau},
  {Razzano}, {Razzaque}, {Reimer}, {Reimer}, {Reposeur}, {Ritz}, {Rochester},
  {Rodriguez}, {Roming}, {Roth}, {Ryde}, {Sadrozinski}, {Sanchez}, {Sander},
  {Saz Parkinson}, {Scargle}, {Schalk}, {Sgr{\`o}}, {Siskind}, {Smith},
  {Spinelli}, {Stamatikos}, {Stecker}, {Stratta}, {Strickman}, {Suson},
  {Swenson}, {Tajima}, {Takahashi}, {Tanaka}, {Thayer}, {Thayer}, {Thompson},
  {Tibaldo}, {Torres}, {Tosti}, {Tramacere}, {Uchiyama}, {Uehara}, {Usher},
  {van der Horst}, {Vasileiou}, {Vilchez}, {Vitale}, {von Kienlin}, {Waite},
  {Wang}, {Wilson-Hodge}, {Winer}, {Wood}, {Yamazaki}, {Ylinen}, \&
  {Ziegler}}]{Abdo090902B}
{Abdo}, A.~A. {et~al.} 2009{\natexlab{a}}, \apjl, 706, L138, 0909.2470

\bibitem[{{Abdo} {et~al.}(2009{\natexlab{b}}){Abdo}, {Ackermann}, {Ajello},
  {Asano}, {Atwood}, {Axelsson}, {Baldini}, {Ballet}, {Barbiellini}, {Baring},
  \& et~al.}]{Abdo090510}
------. 2009{\natexlab{b}}, \nat, 462, 331, 0908.1832

\bibitem[{{Abdo} {et~al.}(2010){Abdo}, {Ackermann}, {Ajello}, {Asano},
  {Atwood}, {Axelsson}, {Baldini}, {Ballet}, {Barbiellini}, {Bastieri},
  {Baughman}, {Bechtol}, {Bellazzini}, {Berenji}, {Bhat}, {Bissaldi},
  {Blandford}, {Bloom}, {Bonamente}, {Borgland}, {Bouvier}, {Bregeon}, {Brez},
  {Briggs}, {Brigida}, {Bruel}, {Burgess}, {Burnett}, {Buson}, {Caliandro},
  {Cameron}, {Caraveo}, {Carrigan}, {Casandjian}, {Cecchi}, {{\c C}elik},
  {Chaplin}, {Charles}, {Chekhtman}, {Chiang}, {Ciprini}, {Claus},
  {Cohen-Tanugi}, {Cominsky}, {Connaughton}, {Conrad}, {Cutini}, {Dermer}, {de
  Angelis}, {de Palma}, {Digel}, {Silva}, {Drell}, {Dubois}, {Dumora},
  {Farnier}, {Favuzzi}, {Fegan}, {Fishman}, {Focke}, {Fortin}, {Frailis},
  {Fukazawa}, {Funk}, {Fusco}, {Gargano}, {Gasparrini}, {Gehrels}, {Germani},
  {Giebels}, {Giglietto}, {Giommi}, {Giordano}, {Glanzman}, {Godfrey},
  {Granot}, {Grenier}, {Grondin}, {Grove}, {Guillemot}, {Guiriec}, {Hanabata},
  {Harding}, {Hayashida}, {Haynes}, {Hays}, {Horan}, {Hughes}, {Jackson},
  {J{\'o}hannesson}, {Johnson}, {Johnson}, {Kamae}, {Katagiri}, {Kataoka},
  {Kawai}, {Kerr}, {Kippen}, {Kn{\"o}dlseder}, {Kocevski}, {Kocian}, {Komin},
  {Kouveliotou}, {Kuehn}, {Kuss}, {Lande}, {Latronico}, {Lemoine-Goumard},
  {Longo}, {Loparco}, {Lott}, {Lovellette}, {Lubrano}, {Madejski}, {Makeev},
  {Mazziotta}, {McBreen}, {McEnery}, {McGlynn}, {Meegan}, {M{\'e}sz{\'a}ros},
  {Meurer}, {Michelson}, {Mitthumsiri}, {Mizuno}, {Moiseev}, {Monte},
  {Monzani}, {Moretti}, {Morselli}, {Moskalenko}, {Murgia}, {Nakamori},
  {Nolan}, {Norris}, {Nuss}, {Ohno}, {Ohsugi}, {Omodei}, {Orlando}, {Ormes},
  {Paciesas}, {Paneque}, {Panetta}, {Parent}, {Pelassa}, {Pepe},
  {Pesce-Rollins}, {Piron}, {Porter}, {Preece}, {Rain{\`o}}, {Rando},
  {Razzano}, {Razzaque}, {Reimer}, {Reimer}, {Reposeur}, {Ripken}, {Ritz},
  {Rochester}, {Rodriguez}, {Roth}, {Ryde}, {Sadrozinski}, {Sanchez}, {Sander},
  {Saz Parkinson}, {Scargle}, {Schalk}, {Sgr{\`o}}, {Siskind}, {Smith},
  {Smith}, {Spandre}, {Spinelli}, {Stamatikos}, {Strickman}, {Suson},
  {Tagliaferri}, {Tajima}, {Takahashi}, {Tanaka}, {Thayer}, {Thayer},
  {Thompson}, {Tibaldo}, {Toma}, {Torres}, {Tosti}, {Tramacere}, {Troja},
  {Uchiyama}, {Usher}, {van der Horst}, {Vasileiou}, {Vilchez}, {Vitale}, {von
  Kienlin}, {Waite}, {Wang}, {Wilson-Hodge}, {Winer}, {Wood}, {Wu}, {Yamazaki},
  {Ylinen}, \& {Ziegler}}]{Abdo081024B}
------. 2010, \apj, 712, 558

\bibitem[{{Abdo} {et~al.}(2009{\natexlab{c}}){Abdo}, {Ackermann}, {Arimoto},
  {Asano}, {Atwood}, {Axelsson}, {Baldini}, {Ballet}, {Band}, {Barbiellini}, \&
  et~al.}]{Abdo080916C}
------. 2009{\natexlab{c}}, Science, 323, 1688

\bibitem[{{Abdo} {et~al.}(2009{\natexlab{d}}){Abdo}, {Ackermann}, {Asano},
  {Atwood}, {Axelsson}, {Baldini}, {Ballet}, {Band}, {Barbiellini}, {Bastieri},
  {Bechtol}, {Bellazzini}, {Berenji}, {Bhat}, {Bissaldi}, {Bloom}, {Bonamente},
  {Borgland}, {Bouvier}, {Bregeon}, {Brez}, {Briggs}, {Brigida}, {Bruel},
  {Burnett}, {Caliandro}, {Cameron}, {Caraveo}, {Casandjian}, {Cecchi},
  {Chaplin}, {Chekhtman}, {Cheung}, {Chiang}, {Ciprini}, {Claus},
  {Cohen-Tanugi}, {Cominsky}, {Connaughton}, {Conrad}, {Cutini}, {Dermer}, {de
  Angelis}, {de Palma}, {Digel}, {Silva}, {Drell}, {Dubois}, {Dumora},
  {Farnier}, {Favuzzi}, {Focke}, {Frailis}, {Fukazawa}, {Fusco}, {Gargano},
  {Gasparrini}, {Gehrels}, {Germani}, {Gibby}, {Giebels}, {Giglietto},
  {Giordano}, {Glanzman}, {Godfrey}, {Goldstein}, {Granot}, {Grenier},
  {Grondin}, {Grove}, {Guillemot}, {Guiriec}, {Hanabata}, {Harding},
  {Hayashida}, {Hays}, {Hughes}, {J{\'o}hannesson}, {Johnson}, {Johnson},
  {Kamae}, {Katagiri}, {Kataoka}, {Kawai}, {Kerr}, {Kn{\"o}dlseder},
  {Kocevski}, {Komin}, {Kouveliotou}, {Kuehn}, {Kuss}, {Latronico}, {Longo},
  {Loparco}, {Lott}, {Lovellette}, {Lubrano}, {Makeev}, {Mazziotta}, {McBreen},
  {McEnery}, {McGlynn}, {Meegan}, {Meurer}, {Michelson}, {Mitthumsiri},
  {Mizuno}, {Monte}, {Monzani}, {Moretti}, {Morselli}, {Moskalenko}, {Murgia},
  {Nakamori}, {Nolan}, {Norris}, {Nuss}, {Ohno}, {Ohsugi}, {Omodei}, {Orlando},
  {Ormes}, {Ozaki}, {Paciesas}, {Paneque}, {Panetta}, {Parent}, {Pelassa},
  {Pepe}, {Pesce-Rollins}, {Piron}, {Porter}, {Preece}, {Rain{\`o}}, {Rando},
  {Razzano}, {Razzaque}, {Reimer}, {Reposeur}, {Ritz}, {Rochester},
  {Rodriguez}, {Roth}, {Ryde}, {Sadrozinski}, {Sanchez}, {Sander}, {Saz
  Parkinson}, {Scargle}, {Sgr{\`o}}, {Siskind}, {Smith}, {Smith}, {Spandre},
  {Spinelli}, {Stamatikos}, {Strickman}, {Suson}, {Tajima}, {Takahashi},
  {Tanaka}, {Thayer}, {Thayer}, {Tibaldo}, {Torres}, {Tosti}, {Tramacere},
  {Uchiyama}, {Usher}, {van der Horst}, {Vasileiou}, {Vilchez}, {Vitale}, {von
  Kienlin}, {Waite}, {Wang}, {Wilson-Hodge}, {Winer}, {Wood}, {Ylinen}, \&
  {Ziegler}}]{Abdo080825C}
------. 2009{\natexlab{d}}, \apj, 707, 580, 0910.4192

\bibitem[{{Achterberg} {et~al.}(2001){Achterberg}, {Gallant}, {Kirk}, \&
  {Guthmann}}]{Achterberg2001}
{Achterberg}, A., {Gallant}, Y.~A., {Kirk}, J.~G., \& {Guthmann}, A.~W. 2001,
  \mnras, 328, 393, arXiv:astro-ph/0107530

\bibitem[{{Ackermann} {et~al.}(2010{\natexlab{a}}){Ackermann}, {Ajello},
  {Baldini}, {Ballet}, {Barbiellini}, {Baring}, {Bastieri}, {Bechtol},
  {Bellazzini}, {Berenji}, {Bhat}, {Bissaldi}, {Blandford}, {Bonamente},
  {Borgland}, {Bouvier}, {Bregeon}, {Brez}, {Briggs}, {Brigida}, {Bruel},
  {Buehler}, {Buson}, {Caliandro}, {Cameron}, {Caraveo}, {Carrigan},
  {Casandjian}, {Cecchi}, {{\c C}elik}, {Charles}, {Chekhtman}, {Chiang},
  {Ciprini}, {Claus}, {Cohen-Tanugi}, {Connaughton}, {Conrad}, {Cutini},
  {Dermer}, {de Angelis}, {de Palma}, {Digel}, {Silva}, {Drell}, {Dubois},
  {Favuzzi}, {Fegan}, {Ferrara}, {Frailis}, {Fukazawa}, {Fusco}, {Gargano},
  {Gasparrini}, {Gehrels}, {Germani}, {Giglietto}, {Giommi}, {Giordano},
  {Giroletti}, {Glanzman}, {Godfrey}, {Granot}, {Grenier}, {Grove},
  {Guillemot}, {Guiriec}, {Hadasch}, {Hays}, {Horan}, {Hughes},
  {J{\'o}hannesson}, {Johnson}, {Johnson}, {Kamae}, {Katagiri}, {Kippen},
  {Kn{\"o}dlseder}, {Kocevski}, {Kuss}, {Lande}, {Latronico}, {Lee}, {Llena
  Garde}, {Longo}, {Loparco}, {Lovellette}, {Lubrano}, {Makeev}, {Mazziotta},
  {McBreen}, {McEnery}, {McGlynn}, {Meegan}, {Mehault}, {M{\'e}sz{\'a}ros},
  {Michelson}, {Mizuno}, {Moiseev}, {Monte}, {Monzani}, {Moretti}, {Morselli},
  {Moskalenko}, {Murgia}, {Nakajima}, {Nakamori}, {Naumann-Godo}, {Nolan},
  {Norris}, {Nuss}, {Ohno}, {Ohsugi}, {Okumura}, {Omodei}, {Orlando}, {Ormes},
  {Ozaki}, {Paciesas}, {Paneque}, {Panetta}, {Parent}, {Pelassa}, {Pepe},
  {Pesce-Rollins}, {Petrosian}, {Piron}, {Porter}, {Preece}, {Racusin},
  {Rain{\`o}}, {Rando}, {Rau}, {Razzano}, {Razzaque}, {Reimer}, {Reimer},
  {Ripken}, {Roth}, {Ryde}, {Sadrozinski}, {Sander}, {Scargle}, {Schalk},
  {Sgr{\`o}}, {Siskind}, {Smith}, {Spandre}, {Spinelli}, {Stamatikos},
  {Strickman}, {Suson}, {Tajima}, {Takahashi}, {Tanaka}, {Thayer}, {Thayer},
  {Tibaldo}, {Torres}, {Tosti}, {Tramacere}, {Uehara}, {Usher},
  {Vandenbroucke}, {van der Horst}, {Vasileiou}, {Vilchez}, {Vitale}, {von
  Kienlin}, {Waite}, {Wang}, {Wilson-Hodge}, {Winer}, {Wood}, {Wu}, {Yamazaki},
  {Yang}, {Ylinen}, {Ziegler}, {Fermi-LAT Collaboration}, \& {Fermi-GBM
  Collaboration}}]{Ackermann090217A}
{Ackermann}, M. {et~al.} 2010{\natexlab{a}}, \apjl, 717, L127, 1007.3409

\bibitem[{{Ackermann} {et~al.}(2010{\natexlab{b}}){Ackermann}, {Asano},
  {Atwood}, {Axelsson}, {Baldini}, {Ballet}, {Barbiellini}, {Baring},
  {Bastieri}, {Bechtol}, {Bellazzini}, {Berenji}, {Bhat}, {Bissaldi},
  {Blandford}, {Bloom}, {Bonamente}, {Borgland}, {Bouvier}, {Bregeon}, {Brez},
  {Briggs}, {Brigida}, {Bruel}, {Buson}, {Caliandro}, {Cameron}, {Caraveo},
  {Carrigan}, {Casandjian}, {Cecchi}, {{\c C}elik}, {Charles}, {Chiang},
  {Ciprini}, {Claus}, {Cohen-Tanugi}, {Connaughton}, {Conrad}, {Dermer}, {de
  Palma}, {Dingus}, {Silva}, {Drell}, {Dubois}, {Dumora}, {Farnier}, {Favuzzi},
  {Fegan}, {Finke}, {Focke}, {Frailis}, {Fukazawa}, {Fusco}, {Gargano},
  {Gasparrini}, {Gehrels}, {Germani}, {Giglietto}, {Giordano}, {Glanzman},
  {Godfrey}, {Granot}, {Grenier}, {Grondin}, {Grove}, {Guiriec}, {Hadasch},
  {Harding}, {Hays}, {Horan}, {Hughes}, {J{\'o}hannesson}, {Johnson}, {Kamae},
  {Katagiri}, {Kataoka}, {Kawai}, {Kippen}, {Kn{\"o}dlseder}, {Kocevski},
  {Kouveliotou}, {Kuss}, {Lande}, {Latronico}, {Lemoine-Goumard}, {Llena
  Garde}, {Longo}, {Loparco}, {Lott}, {Lovellette}, {Lubrano}, {Makeev},
  {Mazziotta}, {McEnery}, {McGlynn}, {Meegan}, {M{\'e}sz{\'a}ros}, {Michelson},
  {Mitthumsiri}, {Mizuno}, {Moiseev}, {Monte}, {Monzani}, {Moretti},
  {Morselli}, {Moskalenko}, {Murgia}, {Nakajima}, {Nakamori}, {Nolan},
  {Norris}, {Nuss}, {Ohno}, {Ohsugi}, {Omodei}, {Orlando}, {Ormes}, {Ozaki},
  {Paciesas}, {Paneque}, {Panetta}, {Parent}, {Pelassa}, {Pepe},
  {Pesce-Rollins}, {Piron}, {Preece}, {Rain{\`o}}, {Rando}, {Razzano},
  {Razzaque}, {Reimer}, {Ritz}, {Rodriguez}, {Roth}, {Ryde}, {Sadrozinski},
  {Sander}, {Scargle}, {Schalk}, {Sgr{\`o}}, {Siskind}, {Smith}, {Spandre},
  {Spinelli}, {Stamatikos}, {Stecker}, {Strickman}, {Suson}, {Tajima},
  {Takahashi}, {Takahashi}, {Tanaka}, {Thayer}, {Thayer}, {Thompson},
  {Tibaldo}, {Toma}, {Torres}, {Tosti}, {Tramacere}, {Uchiyama}, {Uehara},
  {Usher}, {van der Horst}, {Vasileiou}, {Vilchez}, {Vitale}, {von Kienlin},
  {Waite}, {Wang}, {Wilson-Hodge}, {Winer}, {Wu}, {Yamazaki}, {Yang}, {Ylinen},
  \& {Ziegler}}]{Ackermann090510}
------. 2010{\natexlab{b}}, \apj, 716, 1178

\bibitem[{{Asano} {et~al.}(2009){Asano}, {Guiriec}, \&
  {M{\'e}sz{\'a}ros}}]{Asano2009}
{Asano}, K., {Guiriec}, S., \& {M{\'e}sz{\'a}ros}, P. 2009, \apjl, 705, L191,
  0909.0306

\bibitem[{{Belczynski} {et~al.}(2006){Belczynski}, {Perna}, {Bulik},
  {Kalogera}, {Ivanova}, \& {Lamb}}]{Belczynski2006}
{Belczynski}, K., {Perna}, R., {Bulik}, T., {Kalogera}, V., {Ivanova}, N., \&
  {Lamb}, D.~Q. 2006, \apj, 648, 1110, arXiv:astro-ph/0601458

\bibitem[{{Blandford} \& {McKee}(1976)}]{Blandford1976}
{Blandford}, R.~D., \& {McKee}, C.~F. 1976, Physics of Fluids, 19, 1130

\bibitem[{{Corsi} {et~al.}(2010){Corsi}, {Guetta}, \& {Piro}}]{corsi2010}
{Corsi}, A., {Guetta}, D., \& {Piro}, L. 2010, \apj, 720, 1008, 0911.4453

\bibitem[{{De Pasquale} {et~al.}(2010){De Pasquale}, {Schady}, {Kuin}, {Page},
  {Curran}, {Zane}, {Oates}, {Holland}, {Breeveld}, {Hoversten}, {Chincarini},
  {Grupe}, {Abdo}, {Ackermann}, {Ajello}, {Axelsson}, {Baldini}, {Ballet},
  {Barbiellini}, {Baring}, {Bastieri}, {Bechtol}, {Bellazzini}, {Berenji},
  {Bissaldi}, {Blandford}, {Bloom}, {Bonamente}, {Borgland}, {Bouvier},
  {Bregeon}, {Brez}, {Briggs}, {Brigida}, {Bruel}, {Burnett}, {Buson},
  {Caliandro}, {Cameron}, {Caraveo}, {Carrigan}, {Casandjian}, {Cecchi}, {{\c
  C}elik}, {Chekhtman}, {Chiang}, {Ciprini}, {Claus}, {Cohen-Tanugi},
  {Connaughton}, {Conrad}, {Dermer}, {de Angelis}, {de Palma}, {Dingus},
  {Silva}, {Drell}, {Dubois}, {Dumora}, {Farnier}, {Favuzzi}, {Fegan},
  {Fishman}, {Focke}, {Frailis}, {Fukazawa}, {Funk}, {Fusco}, {Gargano},
  {Gasparrini}, {Gehrels}, {Germani}, {Giglietto}, {Giordano}, {Glanzman},
  {Godfrey}, {Granot}, {Greiner}, {Grenier}, {Grove}, {Guillemot}, {Guiriec},
  {Harding}, {Hayashida}, {Hays}, {Horan}, {Hughes}, {Jackson},
  {J{\'o}hannesson}, {Johnson}, {Johnson}, {Kamae}, {Katagiri}, {Kataoka},
  {Kawai}, {Kerr}, {Kippen}, {Kn{\"o}dlseder}, {Kocevski}, {Kuss}, {Lande},
  {Latronico}, {Lemoine-Goumard}, {Longo}, {Loparco}, {Lott}, {Lovellette},
  {Lubrano}, {Makeev}, {Mazziotta}, {McEnery}, {McGlynn}, {Meegan},
  {M{\'e}sz{\'a}ros}, {Meurer}, {Michelson}, {Mitthumsiri}, {Mizuno}, {Monte},
  {Monzani}, {Moretti}, {Morselli}, {Moskalenko}, {Murgia}, {Nolan}, {Norris},
  {Nuss}, {Ohno}, {Ohsugi}, {Omodei}, {Orlando}, {Ormes}, {Paciesas},
  {Paneque}, {Panetta}, {Parent}, {Pelassa}, {Pepe}, {Pesce-Rollins}, {Piron},
  {Porter}, {Preece}, {Rain{\`o}}, {Rando}, {Razzano}, {Reimer}, {Reimer},
  {Reposeur}, {Ritz}, {Rochester}, {Rodriguez}, {Roth}, {Ryde}, {Sadrozinski},
  {Sander}, {Saz Parkinson}, {Scargle}, {Schalk}, {Sgr{\`o}}, {Siskind},
  {Smith}, {Spandre}, {Spinelli}, {Stamatikos}, {Starck}, {Stecker},
  {Strickman}, {Suson}, {Tajima}, {Takahashi}, {Tanaka}, {Thayer}, {Thayer},
  {Thompson}, {Tibaldo}, {Toma}, {Torres}, {Tosti}, {Tramacere}, {Uchiyama},
  {Uehara}, {Usher}, {van der Horst}, {Vasileiou}, {Vilchez}, {Vitale}, {von
  Kienlin}, {Waite}, {Wang}, {Winer}, {Wood}, {Wu}, {Yamazaki}, {Ylinen}, \&
  {Ziegler}}]{DePasquale2010}
{De Pasquale}, M. {et~al.} 2010, \apjl, 709, L146, 0910.1629

\bibitem[{{Dermer}(2004)}]{Dermer2004}
{Dermer}, C.~D. 2004, \apj, 614, 284, arXiv:astro-ph/0403508

\bibitem[{{Fermi-LAT} \& {Fermi-GBM collaborations}(2011)}]{Ackermann090926A}
{Fermi-LAT}, \& {Fermi-GBM collaborations}. 2011, ArXiv e-prints, 1101.2082

\bibitem[{{Freedman} \& {Waxman}(2001)}]{Freedman2001}
{Freedman}, D.~L., \& {Waxman}, E. 2001, \apj, 547, 922, arXiv:astro-ph/9912214

\bibitem[{{Gao} {et~al.}(2009){Gao}, {Mao}, {Xu}, \& {Fan}}]{Gao2009}
{Gao}, W., {Mao}, J., {Xu}, D., \& {Fan}, Y. 2009, \apjl, 706, L33, 0908.3975

\bibitem[{{Ghirlanda} {et~al.}(2010){Ghirlanda}, {Ghisellini}, \&
  {Nava}}]{Ghirlanda2010}
{Ghirlanda}, G., {Ghisellini}, G., \& {Nava}, L. 2010, \aap, 510, L7+,
  0909.0016

\bibitem[{{Ghisellini} {et~al.}(2010){Ghisellini}, {Ghirlanda}, {Nava}, \&
  {Celotti}}]{Ghisellini2010}
{Ghisellini}, G., {Ghirlanda}, G., {Nava}, L., \& {Celotti}, A. 2010, \mnras,
  403, 926, 0910.2459

\bibitem[{{Golenetskii} {et~al.}(2009){Golenetskii}, {Aptekar}, {Mazets},
  {Pal'Shin}, {Frederiks}, {Oleynik}, {Ulanov}, {Svinkin}, \&
  {Cline}}]{Golenetskii2009}
{Golenetskii}, S. {et~al.} 2009, GRB Coordinates Network, 9344, 1

\bibitem[{{Granot} {et~al.}(2010){Granot}, {for the Fermi LAT Collaboration},
  \& {the GBM Collaboration}}]{Granot2010}
{Granot}, J., {for the Fermi LAT Collaboration}, \& {the GBM Collaboration}.
  2010, ArXiv e-prints, 1003.2452

\bibitem[{{Granot} {et~al.}(1999){Granot}, {Piran}, \& {Sari}}]{Granot1999}
{Granot}, J., {Piran}, T., \& {Sari}, R. 1999, \apj, 513, 679,
  arXiv:astro-ph/9806192

\bibitem[{{Granot} \& {Sari}(2002)}]{Granot2002}
{Granot}, J., \& {Sari}, R. 2002, \apj, 568, 820, arXiv:astro-ph/0108027

\bibitem[{{Guiriec} {et~al.}(2009){Guiriec}, {Connaughton}, \&
  {Briggs}}]{Guiriec2009}
{Guiriec}, S., {Connaughton}, V., \& {Briggs}, M. 2009, GRB Coordinates
  Network, 9336, 1

\bibitem[{{Hoversten} {et~al.}(2009){Hoversten}, {Krimm}, {Grupe}, {Kuin},
  {Barthelmy}, {Burrows}, {Roming}, \& {Gehrels}}]{Hoversten2009}
{Hoversten}, E.~A., {Krimm}, H.~A., {Grupe}, D., {Kuin}, N.~P.~M., {Barthelmy},
  S.~D., {Burrows}, D.~N., {Roming}, P., \& {Gehrels}, N. 2009, GCN Report,
  218, 1

\bibitem[{{Huang} {et~al.}(2000{\natexlab{a}}){Huang}, {Dai}, \&
  {Lu}}]{Huang2000mn}
{Huang}, Y.~F., {Dai}, Z.~G., \& {Lu}, T. 2000{\natexlab{a}}, \mnras, 316, 943,
  arXiv:astro-ph/0005549

\bibitem[{{Huang} {et~al.}(2000{\natexlab{b}}){Huang}, {Gou}, {Dai}, \&
  {Lu}}]{Huang2000apj}
{Huang}, Y.~F., {Gou}, L.~J., {Dai}, Z.~G., \& {Lu}, T. 2000{\natexlab{b}},
  \apj, 543, 90, arXiv:astro-ph/9910493

\bibitem[{{Ioka}(2010)}]{Ioka2010}
{Ioka}, K. 2010, Progress of Theoretical Physics, 124, 667, 1006.3073

\bibitem[{{Ioka} \& {Nakamura}(2001)}]{Ioka2001}
{Ioka}, K., \& {Nakamura}, T. 2001, \apjl, 554, L163, arXiv:astro-ph/0105321

\bibitem[{{Kobayashi} {et~al.}(2007){Kobayashi}, {Zhang}, {M{\'e}sz{\'a}ros},
  \& {Burrows}}]{Kobayashi2007}
{Kobayashi}, S., {Zhang}, B., {M{\'e}sz{\'a}ros}, P., \& {Burrows}, D. 2007,
  \apj, 655, 391, arXiv:astro-ph/0506157

\bibitem[{{Kumar} \& {Barniol Duran}(2009)}]{Kumar2009080916C}
{Kumar}, P., \& {Barniol Duran}, R. 2009, \mnras, 400, L75, 0905.2417

\bibitem[{{Kumar} \& {Barniol Duran}(2010)}]{Kumar20093GRBs}
------. 2010, \mnras, 409, 226, 0910.5726

\bibitem[{{Kumar} \& {Panaitescu}(2000)}]{Kumar2000}
{Kumar}, P., \& {Panaitescu}, A. 2000, \apjl, 541, L51, arXiv:astro-ph/0006317

\bibitem[{{Liu} \& {Wang}(2010)}]{Liu2010}
{Liu}, R., \& {Wang}, X. 2010, ArXiv e-prints, 1009.1289

\bibitem[{{Longo} {et~al.}(2009){Longo}, {Moretti}, {Barbiellini}, {Vallazza},
  {Giuliani}, {Cutini}, {Pittori}, {Marisaldi}, {Bulgarelli}, {Gianotti},
  {Trifoglio}, {Di Cocco}, {Labanti}, {Fuschino}, {Galli}, {Chen},
  {Mereghetti}, {Perotti}, {Caraveo}, {Evangelista}, {Del}, {Feroci},
  {Donnarumma}, {Pacciani}, {Soffitta}, {Costa}, {Lazzarotto}, {Lapshov},
  {Rapisarda}, {Pellizzoni}, {Pilia}, {Vercellone}, {Tavani}, {Pucella},
  {D'Ammando}, {Vittorini}, {Argan}, {Trois}, {Piano}, {Sabatini}, {Picozza},
  {Morselli}, {Prest}, {Lipari}, {Zanello}, {Rappoldi}, {Cattaneo}, {Giommi},
  {Santolamazza}, {Verrecchia}, \& {Salotti}}]{Longo2009}
{Longo}, F. {et~al.} 2009, GRB Coordinates Network, 9343, 1

\bibitem[{{Maxham} {et~al.}(2011){Maxham}, {Zhang}, \& {Zhang}}]{Maxham2011}
{Maxham}, A., {Zhang}, B., \& {Zhang}, B. 2011, ArXiv e-prints, 1101.0144

\bibitem[{{McBreen} {et~al.}(2010){McBreen}, {Kr{\"u}hler}, {Rau}, {Greiner},
  {Kann}, {Savaglio}, {Afonso}, {Clemens}, {Filgas}, {Klose}, {K{\"u}pc{\"u}
  Yolda{\c s}}, {Olivares E.}, {Rossi}, {Szokoly}, {Updike}, \& {Yolda{\c
  s}}}]{McBreen2010}
{McBreen}, S. {et~al.} 2010, \aap, 516, A71+, 1003.3885

\bibitem[{{Moderski} {et~al.}(2000){Moderski}, {Sikora}, \&
  {Bulik}}]{Moderski2000}
{Moderski}, R., {Sikora}, M., \& {Bulik}, T. 2000, \apj, 529, 151,
  arXiv:astro-ph/9904310

\bibitem[{{Nakar} {et~al.}(2009){Nakar}, {Ando}, \& {Sari}}]{Nakar2009}
{Nakar}, E., {Ando}, S., \& {Sari}, R. 2009, \apj, 703, 675, 0903.2557

\bibitem[{{Neamus}(2010)}]{Neamus2010}
{Neamus}, A. 2010, ArXiv e-prints, 1005.1051

\bibitem[{{Ohmori} {et~al.}(2009){Ohmori}, {Noda}, {Sonoda}, {Yamauchi},
  {Kono}, {Hayashi}, {Daikyuji}, {Nishioka}, {Ohno}, {Suzuki}, {Kokubun},
  {Takahashi}, {Yamaoka}, {Sugita}, {Nakagawa}, {Tamagawa}, {Hong}, {Vasquez},
  {Uehara}, {Hanabata}, {Fukazawa}, {Iwakiri}, {Tashiro}, {Terada}, {Endo},
  {Onda}, {Sugasahara}, {Urata}, {Enoto}, {Nakazawa}, \&
  {Makishima}}]{Ohmori2009}
{Ohmori}, N. {et~al.} 2009, GRB Coordinates Network, 9355, 1

\bibitem[{{Rau} {et~al.}(2009){Rau}, {McBreen}, \& {Kruehler}}]{Rau2009}
{Rau}, A., {McBreen}, S., \& {Kruehler}, T. 2009, GRB Coordinates Network,
  9353, 1

\bibitem[{{Razzaque}(2010)}]{Razzaque2010}
{Razzaque}, S. 2010, \apjl, 724, L109, 1004.3330

\bibitem[{{Rhoads}(1997)}]{Rhoads1997}
{Rhoads}, J.~E. 1997, \apjl, 487, L1+, arXiv:astro-ph/9705163

\bibitem[{{Rhoads}(1999)}]{Rhoads1999}
------. 1999, \apj, 525, 737, arXiv:astro-ph/9903399

\bibitem[{{Sari}(1997)}]{Sari1997dyn}
{Sari}, R. 1997, \apjl, 489, L37+

\bibitem[{{Sari} \& {Esin}(2001)}]{Sari2001}
{Sari}, R., \& {Esin}, A.~A. 2001, \apj, 548, 787, arXiv:astro-ph/0005253

\bibitem[{{Sari} {et~al.}(1998){Sari}, {Piran}, \&
  {Narayan}}]{Sari1998afterglow}
{Sari}, R., {Piran}, T., \& {Narayan}, R. 1998, \apjl, 497, L17+,
  arXiv:astro-ph/9712005

\bibitem[{{Toma} {et~al.}(2009){Toma}, {Wu}, \&
  {M{\'e}sz{\'a}ros}}]{Toma2009cocoon}
{Toma}, K., {Wu}, X., \& {M{\'e}sz{\'a}ros}, P. 2009, \apj, 707, 1404,
  0905.1697

\bibitem[{{Toma} {et~al.}(2010){Toma}, {Wu}, \& {Meszaros}}]{Toma2010}
{Toma}, K., {Wu}, X., \& {Meszaros}, P. 2010, ArXiv e-prints, 1002.2634

\bibitem[{{Ukwatta} {et~al.}(2009){Ukwatta}, {Barthelmy}, {Baumgartner},
  {Cummings}, {Fenimore}, {Gehrels}, {Hoversten}, {Krimm}, {Markwardt},
  {Palmer}, {Parsons}, {Sakamoto}, {Sato}, {Stamatikos}, \&
  {Tueller}}]{Ukwatta2009}
{Ukwatta}, T.~N. {et~al.} 2009, GRB Coordinates Network, 9337, 1

\bibitem[{{van Eerten} {et~al.}(2010){van Eerten}, {Zhang}, \&
  {MacFadyen}}]{VanEerten2010}
{van Eerten}, H., {Zhang}, W., \& {MacFadyen}, A. 2010, \apj, 722, 235,
  1006.5125

\bibitem[{{Wang} {et~al.}(2010){Wang}, {He}, {Li}, {Wu}, \& {Dai}}]{Wang2010}
{Wang}, X., {He}, H., {Li}, Z., {Wu}, X., \& {Dai}, Z. 2010, \apj, 712, 1232,
  0911.4189

\bibitem[{{Wang} {et~al.}(2006){Wang}, {Li}, \& {M{\'e}sz{\'a}ros}}]{Wang2006}
{Wang}, X., {Li}, Z., \& {M{\'e}sz{\'a}ros}, P. 2006, \apjl, 641, L89,
  arXiv:astro-ph/0601229

\bibitem[{{Wang} {et~al.}(2001){Wang}, {Dai}, \& {Lu}}]{Wang2001}
{Wang}, X.~Y., {Dai}, Z.~G., \& {Lu}, T. 2001, \apjl, 546, L33,
  arXiv:astro-ph/0010320

\bibitem[{{Woods} \& {Loeb}(1999)}]{Woods1999}
{Woods}, E., \& {Loeb}, A. 1999, \apj, 523, 187, arXiv:astro-ph/9903377

\bibitem[{{Yamazaki} {et~al.}(2003){Yamazaki}, {Yonetoku}, \&
  {Nakamura}}]{Yamazaki2003}
{Yamazaki}, R., {Yonetoku}, D., \& {Nakamura}, T. 2003, \apjl, 594, L79,
  arXiv:astro-ph/0306615

\bibitem[{{Zhang} {et~al.}(2003){Zhang}, {Kobayashi}, \&
  {M{\'e}sz{\'a}ros}}]{Zhang2003}
{Zhang}, B., {Kobayashi}, S., \& {M{\'e}sz{\'a}ros}, P. 2003, \apj, 595, 950,
  arXiv:astro-ph/0302525

\bibitem[{{Zhang} {et~al.}(2010){Zhang}, {Zhang}, {Liang}, {Fan}, {Wu},
  {Pe'er}, {Maxham}, {Gao}, \& {Dong}}]{Zhang2010catalog}
{Zhang}, B. {et~al.} 2010, ArXiv e-prints, 1009.3338

\end{thebibliography}
\end{document}